\documentclass[10pt,journal,twoside,compsoc]{IEEEtran}

\usepackage{mathtools}

\usepackage{float}

\PassOptionsToPackage{bookmarks=false}{hyperref}

\usepackage[colorlinks,linkcolor=black,anchorcolor=black,citecolor=black, urlcolor=black]{hyperref}

\usepackage[pdftex]{graphicx}
\graphicspath{{./graphics/}}
\usepackage{subfigure}
\usepackage{multicol}
\usepackage{paralist}
\usepackage{bm}
\usepackage{amsfonts}
\usepackage{url}
\usepackage{multirow}
\usepackage{makecell}
\usepackage{color}

\makeatletter
\newif\if@restonecol
\makeatother

\usepackage[linesnumbered,ruled]{algorithm2e}
\usepackage{algpseudocode}
\usepackage{amsmath}

\usepackage{ragged2e}

\usepackage{fancyhdr} 

\ifCLASSOPTIONcompsoc
  \usepackage[nocompress]{cite}
\else
  \usepackage{cite}
\fi
\ifCLASSINFOpdf

\else

\fi

\newcommand{\tabincell}[2]{\begin{tabular}{@{}#1@{}}#2\end{tabular}}

\begin{document}
%

                                                   
\title{\centering Mobility-aware Seamless Service Migration and Resource Allocation in Multi-edge IoV Systems}


%

\author{Zheyi Chen,~\IEEEmembership{Member,~IEEE},
	Sijin Huang,
	Geyong Min,~\IEEEmembership{Member,~IEEE},
        Zhaolong Ning,~\IEEEmembership{Senior Member,~IEEE},
	Jie Li,~\IEEEmembership{Fellow, IEEE},
        and Yan Zhang,~\IEEEmembership{Fellow, IEEE} 


\IEEEcompsocitemizethanks{

    \IEEEcompsocthanksitem Zheyi Chen and Sijin Huang are with the College of Computer and Data Science, Fuzhou University, Fuzhou 350116, China, the Engineering Research Center of Big Data Intelligence, Ministry of Education, Fuzhou 350002, China, and also with the Fujian Key Laboratory of Network Computing and Intelligent Information Processing (Fuzhou University), Fuzhou 350116, China. E-mail: z.chen@fzu.edu.cn, sjinhuang@163.com.

    \IEEEcompsocthanksitem Geyong Min is with the Department of Computer Science, Faculty of Environment, Science and Economy, University of Exeter, Exeter EX4 4QF, United Kingdom. E-mail: g.min@exeter.ac.uk.

    \IEEEcompsocthanksitem Zhaolong Ning is with the School of Communications and Information Engineering, Chongqing University of Posts and Telecommunications, Chongqing 400065, China. E-mail: z.ning@ieee.org.

    \IEEEcompsocthanksitem Jie Li is with the Department of Computer Science and Engineering, Shanghai Jiao Tong University, Shanghai 200240, China. E-mail: lijiecs@sjtu.edu.cn.

    \IEEEcompsocthanksitem Yan Zhang is with the Department of Informatics, University of Oslo, 0316 Oslo, Norway. E-mail: yanzhang@ieee.org.

}

\thanks{
	Corresponding authors: Geyong Min and Zhaolong Ning.}
}

%
%

\markboth{IEEE TRANSACTIONS ON MOBILE COMPUTING,~Vol.~xx, No.~xx, XX~2024}%
{Chen \MakeLowercase{\textit{et al.}}: Mobility-aware Seamless Service Migration and Resource Allocation in Multi-edge IoV Systems}

%

\IEEEtitleabstractindextext{%
\begin{abstract}
\justifying
Mobile Edge Computing (MEC) offers low-latency and high-bandwidth support for Internet-of-Vehicles (IoV) applications. However, due to high vehicle mobility and finite communication coverage of base stations, it is hard to maintain uninterrupted and high-quality services without proper service migration among MEC servers. Existing solutions commonly rely on prior knowledge and rarely consider efficient resource allocation during the service migration process, making it hard to reach optimal performance in dynamic IoV environments. To address these important challenges, we propose \textit{SR-CL}, a novel mobility-aware seamless Service migration and Resource allocation framework via Convex-optimization-enabled deep reinforcement Learning in multi-edge IoV systems. First, we decouple the Mixed Integer Nonlinear Programming (MINLP) problem of service migration and resource allocation into two sub-problems. Next, we design a new actor-critic-based asynchronous-update deep reinforcement learning method to handle service migration, where the delayed-update actor makes migration decisions and the one-step-update critic evaluates the decisions to guide the policy update. Notably, we theoretically derive the optimal resource allocation with convex optimization for each MEC server, thereby further improving system performance. Using the real-world datasets of vehicle trajectories and testbed, extensive experiments are conducted to verify the effectiveness of the proposed \textit{SR-CL}. Compared to benchmark methods, the \textit{SR-CL} achieves superior convergence and delay performance under various scenarios.

\justifying
\end{abstract}

}

\maketitle


\IEEEdisplaynontitleabstractindextext

%
\IEEEpeerreviewmaketitle

\ifCLASSOPTIONcompsoc
\IEEEraisesectionheading{\section{Introduction}\label{sec:introduction}}
\else
\fi

\IEEEPARstart
{W}{ith} the fast development of 5G techniques, Internet-of-Vehicles (IoV) has become a new module in smart cities. 
Equipped with intelligent sensors and components, vehicles are capable of hosting various applications such as automatic driving, image recognition, and path planning \cite{pliatsios2022joint1}. 
However, the real-time demands of IoV applications pose significant challenges for onboard processors with limited computational capabilities \cite{ning2022partial}. 
Although Cloud Computing (CC) allows offloading tasks to the remote cloud for processing, the excessive delay caused by long distance is inevitable \cite{luo2021software}. 
To relieve this issue, the emerging Mobile Edge Computing (MEC) offers low-latency and high-bandwidth services by deploying resources at the network edge \cite{gao2022an}. 
Through combining MEC and IoV, a new computing architecture has emerged to enable efficient interconnection and real-time data exchange between vehicles and road infrastructures \cite{Luo2024edgecooper}. 
A typical MEC-enabled IoV system usually considers an area covered by multiple Base Stations (BSs), and vehicles can offload their applications’ tasks to nearby MEC servers that are collocated with BSs, where virtualization is deemed as a key technique to conduct resource management \cite{dai2024survey}. 
When vehicles offload tasks, MEC servers create dedicated service instances via virtualization techniques for the vehicles and allocate proper resources to them \cite{zhou2023service}. 
The service instances encapsulate the runtime data and user context information, which offer fine computing services for vehicles while ensuring resource isolation.

Considering the high mobility of vehicles and finite communication coverage of BSs, it is challenging for a single MEC server to provide seamless services to a vehicle without multi-edge cooperation, which may seriously degrade the Quality-of-Service (QoS) \cite{zhang2023trajectory}.
To guarantee high QoS, the service instances created by MEC servers are expected to be properly migrated along with the mobility of vehicles \cite{kim2022modems6}. 
The performance of service migration depends on multiple factors including vehicle mobility, attribute of offloaded tasks, and available MEC resources. 
An improper migration strategy might cause excessive task response delay and degraded QoS. 
Commonly, the process of service migration in multi-edge IoV systems can be regarded as a long-term decision-making problem. 
The current migration decisions might affect future system performance, and thus it is challenging to optimize the long-term performance without foreknowing the potential mobility of vehicles. 
Moreover, there might exist mutual influence among the migration decisions of different vehicles, making it extremely hard to simultaneously optimize service migration for all vehicles with the concern of minimizing system delay.

Following migration decisions, service instances of vehicles will be migrated to target MEC servers for processing tasks. 
Compared to the remote cloud, MEC servers own limited computational resources, and the continuous influx of tasks commonly imposes various resource demands. 
Therefore, it is necessary to allocate suitable resources to service instances on resource-constrained MEC servers. 
Most of the existing studies \cite{velrajan2022qos7, ouyang2018follow8} did not well consider optimizing resource allocation when they dealt with service migration, which seriously hindered the improvement of system performance. 
Few studies \cite{liang2021multi9, liu2023joint10} investigated the joint optimization of service migration and resource allocation, regarded as a Mixed Integer Nonlinear Programming (MINLP) problem. 
These studies typically adopted classic optimization theories, which usually required numerous iterations and caused excessive overheads. 
Meanwhile, when facing complex and dynamic multi-edge IoV systems, they did not well consider future vehicle mobility and thus they easily fell into the local optimum. 
As an emerging technique, Deep Reinforcement Learning (DRL) \cite{arulkumaran2017deep11} is deemed a promising method for handling this problem. 
Through interactive learning with the environment, the DRL agent can gradually adjust the policy to maximize long-term cumulative rewards. 
Most of the existing studies \cite{wang2021delay12, zhang2019task13, peng2019deep14} adopted the value-based DRL, which learned deterministic policies by selecting actions with the maximum Q-value. 
However, the huge decision-making space in multi-edge IoV systems may result in low learning efficiency and even training failure. 
In contrast, the policy-based DRL \cite{lan2020deep15, QIN2024cotm} can handle the large action space by directly outputting the probability distribution of actions, but high variance may happen when estimating the policy gradient. Moreover, they reveal limited capability to explore continuous action space, and thus the policy may easily fall into the local optimal.

To address the above important challenges, we propose \textit{SR-CL}, a novel mobility-aware seamless Service migration and Resource allocation framework via Convex-optimization-enabled deep reinforcement Learning in multi-edge IoV systems. The \textit{SR-CL} introduces an effective problem decoupling to relieve the limitations of classic DRL on exploring the high-dimensional space of continuous actions. By leveraging the convex optimization theory, the optimal resource allocation can be obtained and then embedded into the DRL model, thereby significantly reducing the dimension of the action space in the original problem. Moreover, a new asynchronous update mechanism is developed to further alleviate the training fluctuations when exploring the optimal policy. The main contributions of this work are summarized as follows.
\begin{itemize}
	\item We design a unified model for service migration and resource allocation in complex and dynamic multi-edge IoV systems. First, the long-term QoS is set as the optimization objective that consists of migration, communication, and computation delays. Next, two sub-problems including service migration and resource allocation of the original MINLP problem are decoupled and formulated, respectively.
	\item For the sub-problem of service migration, we propose a new actor-critic-based DRL method with the asynchronous update to explore the optimal policy. Specifically, the actor with a delayed update makes migration decisions based on system states, while the critic with a one-step update evaluates the decisions and offers accurate guidance for updating the actor.
	\item For the sub-problem of resource allocation, we develop a customized convex-optimization-based method. First, we prove that the sub-problem is a convex programming problem by using the Hessian matrix with constraints. Next, based on the convex optimization theory, we define the generalized Lagrange function and Karush-Kuhn-Tucker (KKT) conditions. Finally, we theoretically derive the optimal resource allocation for each MEC server under given migration decisions.
	\item Using the real-world datasets of vehicle trajectories in Rome City and testbed, we conduct extensive experiments to validate the effectiveness of the proposed \textit{SR-CL}. We focus on the area of the city center with multiple complex mobility patterns of vehicles, making the experiments more practical. The results show that the \textit{SR-CL} can always obtain better convergence and delay performance than other benchmark methods under different scenarios.
\end{itemize}

The rest of this paper is organized as follows. Section \ref{sec:related_work} reviews the related work. Section \ref{sec:systemmodel} describes the model and formulates the problem. Section \ref{sec:gru-sa} details the proposed \textit{SR-CL}. Section \ref{sec:experiments} conducts the performance evaluation. Section \ref{sec:conclusion} concludes this paper.

%
%
%
%

\section{Related Work}\label{sec:related_work}
The problems of service migration and resource allocation in MEC have garnered considerable research attention while many scholars have made contributions. In this section, we review and analyze the related studies.

\subsection{Service Migration}
Network interruption, signal attenuation, and service instability may happen when users or vehicles move, leading to degraded QoS. 
To relieve this issue, the emerging technique of service migration is expected to maintain service continuity and data integrity. 
Velrajan $et~al.$ \cite{velrajan2022qos7} designed a closed-loop particle swarm optimization approach for service migration, considering QoS, resource utilization, and application characteristics. 
Ouyang $et~al.$ \cite{ouyang2018follow8} divided the service placement into a series of sub-problems and addressed them via Lyapunov optimization theory under constraints of the long-term budget. 
Liang $et~al.$ \cite{liang2021multi9} relied on an optimal iterative scheme to deal with the integer-relaxed issue of service migration in multi-cell MEC. 
Liu $et~al.$ \cite{liu2023joint10} investigated the joint problem of request assignment and service migration, aiming to minimize the total response delay of requests in MEC. 
Scotece $et~al.$ \cite{scotece2023handling} proposed a data-handoff framework to migrate inference and training tasks between different edge nodes for reducing accuracy degradation.
Maia $et~al.$ \cite{maia2021improved} proposed an improved Genetic Algorithm (GA) to cope with the load distribution and service placement in edge computing. 
Wang $et~al.$ \cite{wang2021delay12} designed a Q-Learning-based approach for solving the online micro-service migration among edge servers. 
Zhang $et~al.$ \cite{zhang2019task13} implemented a single-user service migration strategy based on Deep Q-Network (DQN) to reduce migration costs and ensure QoS. 
Peng $et~al.$ \cite{peng2019deep14} developed a DQN-based service migration method to reconcile QoS and migration overheads. 
Zhang $et~al.$ \cite{zhang2022deep18} proposed a DQN-based service migration approach with the consideration of network status and user mobility. 
Labriji $et~al.$ \cite{Labriji2021mobility} designed a proactive real-time migration method to migrate computing services in vehicular networks, which integrated a mobility prediction algorithm based on neural networks and Markov chains with a Lyapunov-based online optimizer. 
Perin $et~al.$ \cite{Perin2023ease} developed a resource scheduler that incorporated a dual-layer optimization mechanism to reduce the carbon footprint of edge networks while ensuring QoS.

In general, most of the existing studies handled service migration by using classic optimization theories or heuristics. 
The optimization theories usually require numerous iterations, resulting in high system costs. 
The heuristics commonly rely on expert experiences, which limits their applicability and causes excessive rule-setting overheads. 
Moreover, due to the complexity and dynamics of multi-edge IoV environments, the above methods might tend to fall into the local optimum. 
Some studies adopted the value-based DRL for service migration, exhibiting good performance with small action space. 
Nevertheless, as the action space grows explosively, they cannot collect sufficient training samples to learn an accurate value function, causing unstable learning processes and unsatisfying migration results. 
Although few studies \cite{lan2020deep15, QIN2024cotm} can better handle this problem by using the policy-based DRL to directly output the probability distribution of actions, a high variance may occur when estimating the policy gradient. This problem is caused by policy parameterization and environment dynamics, significantly degrading the stability and efficiency of the learning process. Moreover, most of these studies did not conduct further optimization for resource allocation, which seriously restrained the performance enhancement of service migration.

\subsection{Resource Allocation}
In MEC systems, end devices can offload application tasks to nearby MEC servers for processing, alleviating their capacity constraints. 
However, compared to the remote cloud, MEC servers own fewer resources while facing a continuous influx of tasks with various demands, and thus it is necessary to design proper strategies for resource allocation. 
Su $et~al.$ \cite{su2023joint} proposed two computation offloading and resource allocation methods by combining the weighted minimum Mean Squared Error (MSE) algorithm, quadratic transformation, and difference of convex functions algorithm.
Du $et~al.$ \cite{du2018computation19} designed a sub-optimal strategy for computation offloading and resource allocation in hybrid fog/cloud systems. 
Deng $et~al.$ \cite{deng2016optimal20} proposed an approximate optimization method for resource allocation to balance delay and power consumption. 
Li $et~al.$ \cite{li2021cooperative21} investigated the joint problem of computation offloading and resource allocation in multi-edge scenarios and designed a sub-optimal iterative scheme based on alternating and convex optimization, aiming to reduce energy consumption under various constraints. 
Jošilo $et~al.$ \cite{josilo2019wireless22} proposed a decentralized equilibrium calculation approach for resource allocation to reduce task completion delay in edge systems. 
Huang $et~al.$ \cite{huang2020deep23} designed an online DRL-based resource allocation framework to maximize the computation rate with constraints of energy consumption. 
Zhou $et~al.$ \cite{zhou2022deep24} developed an improved DRL-based computation offloading and resource allocation method to reduce the energy consumption of MEC systems with delay constraints. 
Seid $et~al.$ \cite{seid2021collaborative25} designed a model-free DRL-based method for resource allocation to minimize task execution delay and energy consumption in multi-UAV networks. 

Generally, the above studies targeted the problem of resource allocation or computation offloading in MEC, but they did not well consider the mobility of users or vehicles during the decision-making process. In real-world scenarios, the geographic locations of users or vehicles may change frequently, posing significant challenges in maintaining high-quality and seamless services. In this regard, service instances are expected to be dynamically migrated according to the mobility of users or vehicles, leading to time-varying numbers of service instances and fluctuating workloads on different MEC servers. Under such dynamic and complex scenarios, it is extremely challenging to allocate proper resources to service instances on MEC servers, which will severely affect the performance of service migration.

To the best of our knowledge, this is the first of its kind to propose a mobility-aware framework that integrates DRL with convex optimization theory for addressing the joint problem of seamless service migration and resource allocation in multi-edge IoV systems.

\section{System Model and Problem Formulation}\label{sec:systemmodel}

\begin{figure}[!ht]
	\centering
	\begin{center}
		\includegraphics*[width=1.00\linewidth]{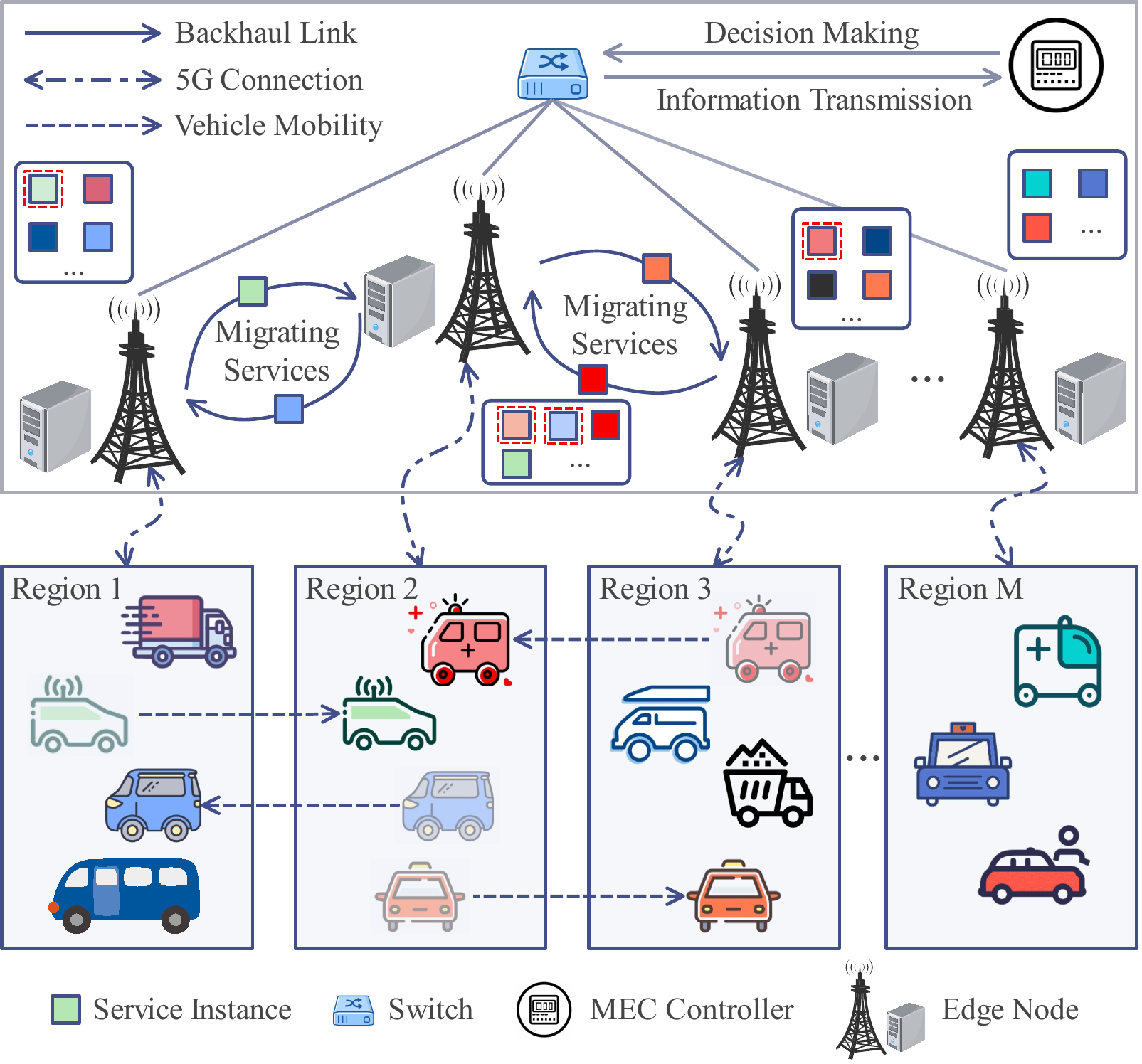}
            \vspace{-0.6cm}
		\caption{The proposed multi-edge IoV system.}
            \vspace{-0.7cm}
		\label{System_model}
	\end{center}
\end{figure}

As shown in Fig. \ref{System_model}, the proposed multi-edge IoV system consists of a MEC controller, $M$ BSs, and $U$ Intelligent Vehicles (IVs). 
Each BS is equipped with a MEC server, named edge node. 
The set of edge nodes is denoted as ${\mathcal F}=\{f_{1},f_{2},...,f_{m},...,f_{M}\}$, and the set of IVs is denoted as ${\mathcal {IV}}=\{IV_{1},IV_{2},...,IV_{u},...IV_{U}\}$. In the proposed system, $\textstyle IV_{u}$ first sends a task to its service instance via the 5G network, and the BS forwards the task to the MEC server that locates the service instance. Next, the MEC server processes the task and returns results. 
During this process, the switch gathers and transmits information to the MEC controller (where the DRL agent is located), which generates migration decisions according to the mobility of IVs, ensuring seamless and high-quality services.

Specifically, the proposed system is running in discrete time slots, and locations of $\textstyle IV_{u}$ may change at the beginning of each time slot $t$, where $t\in\{0,1,2,...,T\}$. Meanwhile, the intelligent application on $\textstyle IV_{u}$ generates a computation-intensive task at each time slot, denoted by $Task_{u,t}$. Due to the limited computational capability of $\textstyle IV_{u}$, these tasks are continuously offloaded to the MEC server for processing. At the first time slot, $\textstyle IV_{u}$ accesses the system via its nearest BS and creates a service instance on the associated MEC server. In subsequent time slots, this service instance offers computing services to $\textstyle IV_{u}$. 
The moving $\textstyle IV_{u}$ connects to the nearest edge node and sends tasks. 
Due to the high mobility of vehicles, $\textstyle IV_{u}$ may drive out of the communication coverage of the current edge node. Thus, the connection will be interrupted and $\textstyle IV_{u}$ will reconnect to the nearest edge node. 
The tasks from IVs are processed by their own service instances, which run in parallel and consume part of the computational resources on MEC servers.

At the time slot$\ t$, the edge node connected to $\textstyle IV_{u}$ is denoted as $f_{i(1\leq i \leq M)}\in{\mathcal{F}}$, and the edge node that locates the service instance of $\textstyle IV_{u}$ is denoted as $f_{j(1\leq j \leq M)}\in{\mathcal{F}}$. Meanwhile, edge nodes are inter-connected via stable backhaul links. In the case that $\textstyle IV_{u}$ disconnects to $f_{j}$, its task can still be transmitted to $f_{j}$ via the backhaul link, but this may cause extra communication delay. This extra delay can be avoided if the service instance moves to $f_{i}$, but the migration of service data also leads to migration delay. Therefore, it is tough to reduce these delays effectively while ensuring QoS by determining the suitable time and destination of service migration. Moreover, when a service instance is migrated to an edge node, proper resources should also be allocated to reduce task computation delay. With the above concerns, we formulate the joint optimization problem of service migration and resource allocation with the objective of reducing system delays including migration, communication, and computation delays.

\subsection{Migration Model}

We define $f_{c(1\leq c \leq M)}\in{\mathcal{F}}$ as the edge node that locates the service instance of $\textstyle IV_{u}$ at the time slot $t$-1 and $x_{u,t}\in\{1,2,...,M\}$ as the migration decision of $\textstyle IV_{u}$ at the time slot $t$, and thus $f_{j}$ is determined by $x_{u,t}$ (i.e., $j=x_{u,t}$).
The set of migration decisions of all IVs at the time slot $t$ is denoted as ${\mathcal{X}_{t} = \{{x_{u,t} | IV_{u} \in \mathcal{IV}\}}}$.  
Moreover, we use ${\delta_{u,t}^{c,j}}$ to measure the hop distance between $f_{c}$ and $f_{j}$. If $c = j$,then $ {\delta_{u,t}^{c,j}} = 0$, indicating that no service migration happens at the current time slot. 
Otherwise, the service instance of $\textstyle IV_{u}$ will be migrated from $f_{c}$ to $f_{j}$, and the migration delay occurs. Such delay is commonly caused by service interruption and grows with the increasing service data amount and hop distance. 
Therefore, the migration delay is a monotonic non-decreasing function with respect to $ {\delta_{u,t}^{c,j}}$, and it is defined as
\begin{equation}
    \label{migration_time}
	MT_{u,t}=\left\{\begin{array}{ll}
		0, & \delta_{u,t}^{c,j}=0 \\
		\frac{S_{u,t}}{\chi}+\mu^{y} \delta_{u,t}^{c,j}, & \delta_{u,t}^{c,j} \neq 0
	\end{array},\right.
\end{equation}
where ${S_{u,t}}$ is the amount of service data, $ {\chi}$ is the bandwidth of the backhaul link, and $\mu^{y}$ is the unit migration delay coefficient that indicates the migration delay per hop. 

\subsection{Communication Model}

After service migration, $\textstyle IV_{u}$ will offload $Task_{u,t}$ to its service instance on $f_{j}$ for processing, and the communication delay occurs, which consists of the data transmission delay between $\textstyle IV_{u}$ and $f_{i}$ and the data transmission delay between $f_{i}$ and $f_{j}$ on the backhaul link. The Signal-to-Noise Ratio (SNR) between $\textstyle IV_{u}$ and $f_{i}$ is defined as
\begin{equation}
	SNR_{u,i}=\frac{P_{u}\alpha}{\sigma^{2}|Len_{u,i}|^{2}},
\end{equation}
where $P_{u}$ is the transmission power of $\textstyle IV_{u}$, $\alpha$ is the channel gain per unit distance, $Len_{u,i}$ is the distance between $\textstyle IV_{u}$ and $f_{i}$, and $\sigma^{2}$ is the Gaussian noise.

Next, the total bandwidth of a BS is denoted as $B$, and the Orthogonal Frequency Division Multiplexing (OFDM) is used to equally distribute $B$ to vehicles at each time slot. Thus, the data transmission delay of $\textstyle IV_{u}$ is defined as
\begin{equation}
	PT_{u,t}=\frac{D_{u,t}}{B_{u,t}\log_{2}(1+SNR_{u,i})},
\end{equation}
where $B_{u,t}$ indicates the available bandwidth of $\textstyle IV_{u}$, and $D_{u,t}$ is the data amount of $Task_{u,t}$.

If $i\neq j$, $Task_{u,t}$ will be transmitted on the backhaul link. The data transmission delay on the backhaul link depends on $D_{u,t}$ and the hop distance between $f_{i}$ and $f_{j}$. Similar to Eq. (\ref{migration_time}), we use $\phi_{u,t}^{i,j}$ to measure the hop distance. Since the data amount of task results is small, the downloading delay is negligible. Thus, the transmission delay on the backhaul link is defined as
\begin{equation}
	ST_{u,t}=\left\{\begin{array}{ll}
		0, & \phi_{u,t}^{i,j}=0 \\
		\frac{D_{u,t}}{\chi}+\mu^{h} \phi_{u,t}^{i,j}, & \phi_{u,t}^{i,j} \neq 0
	\end{array},\right.
\end{equation}
where $\mu^{h}$ is the unit transmission delay coefficient that indicates the transmission delay per hop.

Therefore, the communication delay is defined as
\begin{equation}
	HT_{u,t}=PT_{u,t}+ST_{u,t}.
\end{equation}

\subsection{Computation Model}

When $Task_{u,t}$ is offloaded to $f_{j}$, the MEC server will allocate computational resources to the service instance for processing it. The CPU cycles required to process $Task_{u,t}$ are defined as
\begin{equation}
	K_{u,t}=D_{u,t}C_{u,t},
\end{equation}
where $C_{u,t}$ is the computational density of $Task_{u,t}$ that indicates the CPU cycles for processing one-bit task data.

The maximum computational capability (i.e., CPU frequency) of a MEC server is denoted as $F$, the proportion of computational resources allocated to the service instance of $IV_{u}$ is denoted as $e_{u,t}$, and the set of allocation decisions of all IVs at time slot $t$ is denoted as $\mathcal{E}_{t} = \{e_{u,t} | IV_{u} \in \mathcal{IV}\}$. Therefore, the computation delay is defined as
\begin{equation}
	CT_{u,t}={\frac{K_{u,t}}{e_{u,t}F}}.
\end{equation}

\subsection{Problem Formulation}
Under the decision sets of service migration $\mathcal{X}_{t}$ and resource allocation $\mathcal{E}_{t}$, we define the total delay of the proposed multi-edge IoV system as follows including migration, communication, and computation delays.

\begin{equation}
	\mathcal{G}(\mathcal{X}_{t}, \mathcal{E}_{t}) = \sum_{IV_{u} \in \mathcal{IV}}\left(MT_{u,t}+HT_{u,t}+CT_{u,t}\right).
\end{equation}

Within $T$ time slots, we aim to minimize the long-term system delays and guarantee seamless service provisioning. Thus, the MINLP optimization problem is formulated as
\begin{equation}
	\begin{aligned}
		\textit{P}1: \min _{\mathcal{X}_{t}, \mathcal{E}_{t}} & \sum_{t=0}^{T} \mathcal{G}(\mathcal{X}_{t}, \mathcal{E}_{t}) \\
		\text { s.t. } \quad C 1 & : x_{u,t} \in \{1,2,...,M\}, t \in T, IV_{u} \in \mathcal{IV}, \\
		C 2 & : e_{u,t} \in[0,1], t \in T, IV_{u} \in \mathcal{IV}, \\
		C 3 & : \sum_{IV_{u} \in \mathcal{IV}_{m}} e_{u,t} \leq 1, t \in T, f_{m} \in \mathcal{F} ,
	\end{aligned}
\end{equation}
where the IVs that migrate their service instances to the edge node $f_{m}$ are denoted as the set $\mathcal{IV}_{m}$. 
\textit{C}1 indicates the constraint of a migration decision.  
\textit{C}2 indicates the constraint of the resource proportion allocated to a service instance. 
A service instance can only run on a single edge node in each time slot. \textit{C}3 indicates the constraint of the proportion sum of the computational resources allocated to service instances hosted on the edge node $f_{m}$.

\textbf{Lemma 1.} \textit{P}1 \textit{is an NP-hard problem}.

\textbf{Proof.} For clarity, we introduce the Knapsack Problem (KP) that is proved to be NP-hard. In KP, there is a backpack with the capacity of $W$ and $G$ items, denoted by the set ${\mathcal{B}}=\{b_{1},b_{2},...,b_{g},...,b_{G}\}$. The aim of solving KP is to find an item sub-set $\mathcal{B}' \subseteq\mathcal{B}$ that can maximize the total value of the items in the backpack. Thus, the KP is defined as
\begin{equation}
	\max \sum_{b_{g} \in \mathcal{B}'} v_{g} \quad \quad \text{s.t.} \sum_{b_{g} \in \mathcal{B}'} w_{g} \leq W.
\end{equation}
where $w_{g}$ and $v_{g}$ are the weight and value of item $b_{g}$.

Next, we consider a specific example of \textit{P}1. At the time slot $t$, there are $U$ IVs and $M$ edge nodes. IVs are moving and the computational resources of edge nodes are limited, thus it is necessary to migrate service instances among edge nodes. After service migration, $\textstyle IV_{u}$ can offload its task to the service instance for processing. The total computational resources of MEC servers is denoted as $F_{ALL}$, and the total delay is redefined as $AT_{u,t}$, where $AT_{u,t} = - ( MT_{u,t}+HT_{u,t}+CT_{u,t})$. Therefore, the optimization problem in this example can be described as
\begin{equation}
	\begin{aligned}
			\max _{\mathcal{X}_{t}, \mathcal{E}_{t}} & \sum_{IV_{u} \in \mathcal{IV}} AT_{u,t} = \max _{\mathcal{X}_{t}, \mathcal{E}_{t}} - \mathcal{G}(\mathcal{X}_{t}, \mathcal{E}_{t})  \\
			\text { s.t. }  &  \sum_{IV_{u} \in \mathcal{IV}} e_{u,t}F \leq F_{ALL}.
	\end{aligned}
\end{equation}

Through the above analysis, we prove that this example of \textit{P}1 is a KP and NP-hard. Further, \textit{P}1 considers the long-term system optimization and requires problem stacking of this example over multiple time slots. Therefore, \textit{P}1 is an NP-hard problem.

\subsection{Problem Decoupling}
Notably, service migration and resource allocation belong to two dimensions of \textit{P}1. 
Service migration focuses on service instances within the regional scope, while resource allocation is specific to service instances on a single edge node, posing challenges in finding a unified strategy.
Considering the distinction of decision-making types between service migration and resource allocation,
inspired by \cite{tran2019jointtask}, we have discovered that by fixing the migration decisions $\mathcal{X}_{t}$, \textit{P}1 can be decoupled into multiple sub-problems with separate optimization objectives and constraints. Specifically, we employ the Tammer decoupling method \cite{tammer1987application}, which can promise the optimality of the solution, to decouple the highly complex \textit{P}1 into two sub-problems with lower complexity. First, we rewrite \textit{P}1 as 
\begin{equation}
    \label{rewritep1}
	\operatorname*{min}_{\mathcal{X}_{t}} \sum_{t=0}^{T} \left(\operatorname*{min}_{\mathcal{E}_{t}} \mathcal{G}\left(\mathcal{X}_{t},\mathcal{E}_{t}\right)\right)  \quad \quad \text { s.t. } C1-C3.
\end{equation}

It can be observed that the three constraints in Eq. (\ref{rewritep1}) regarding $\mathcal{X}_{t}$ and $\mathcal{E}_{t}$ are decoupled. Therefore, we can decouple \textit{P}1 into two sub-problems as follows.
\begin{itemize}
	\item \textit{P}2: Minimize long-term system delays by optimizing service migration. This sub-problem is defined as
		\begin{equation}
			\textit{P}2: \min _{\mathcal{X}_{t}}  \sum_{t=0}^{T} \mathcal{G}^{*}\left(\mathcal{X}_{t}\right) \quad \quad \text { s.t. } C1.
			\end{equation}
	
	\item \textit{P}3: Minimize the computation delay under the given migration decisions by optimizing resource allocation. This sub-problem is defined as
		\begin{equation}\label{p3}
			\textit{P}3: \mathcal{G}^{*}\left(\mathcal{X}_{t}\right)  = \min _{\mathcal{E}_{t}} \mathcal{G}\left(\mathcal{X}_{t}, \mathcal{E}_{t}\right) \quad \quad \text { s.t. }  C 2- C3 .
			\end{equation}
	\text{ }\text{ }\text{ }\text{ } Once $\mathcal{X}_{t}$ is determined, $\sum_{IV_{u} \in \mathcal{IV}}\left(MT_{u,t}+HT_{u,t}\right)$ can be obtained subsequently, and we denote it as $\mathcal{I}_{t}$. At this point, we need to optimize $\mathcal{E}_{t}$ for minimizing $\sum_{IV_{u} \in \mathcal{IV}} CT_{u,t}$, which is the objective of \textit{P}3. It should be noted that the resource allocation processes on different MEC servers are independent and parallel. Thus, we can further transform \textit{P}3 into the following form.
	\begin{equation}
		\begin{aligned}
		 \min _{\mathcal{E}_{t}} \mathcal{G}\left(\mathcal{X}_{t}, \mathcal{E}_{t}\right) = & \mathcal{I}_{t} + \min _{\mathcal{E}_{t}}  \sum_{IV_{u} \in \mathcal{IV}} CT_{u,t} \\
		 = &  \mathcal{I}_{t} + \min_{\mathcal{E}_{t}} \sum_{f_{m} \in \mathcal{F}}  \sum_{IV_{u} \in \mathcal{IV}_{m}} CT_{u,t} \\
			\text { s.t. }  \quad C & 2- C3 .
		\end{aligned}
	\end{equation}
\end{itemize}


\section{The Proposed \textit{SR-CL}} \label{sec:gru-sa}
To address \textit{P}2 and \textit{P}3, we propose a novel mobility-aware seamless Service migration and Resource allocation framework via Convex-optimization-enabled deep reinforcement Learning (\textit{SR-CL}) in multi-edge IoV systems. 


\subsection{Service Migration with Improved DRL}
Service migration in multi-edge IoV systems is a sequential decision-making problem that can be modeled as a Markov Decision Process (MDP). In DRL, the 5-tuple $(\mathcal{S}, \mathcal{A}, \mathcal{P}, \mathcal{R}, \gamma)$ is commonly employed to handle the MDP, where $\mathcal{S}$, $\mathcal{A}$, $\mathcal{P}$, $\mathcal{R}$, and $\gamma$ indicate the state space, action space, state transition, reward function, and discount factor, respectively. The policy $\pi(\cdot|s_{t})$ indicates the action distribution at the state $s_{t}$. Given a policy $\pi$, the DRL agent first chooses and executes an action $a_{t}$ at the state $s_{t}$. Next, the environment feedbacks the instant reward $r_{t}$ and steps to the next state $s_{t+1}$. Through this interaction process, the DRL agent will obtain a trajectory $\tau=\{s_{t},a_{t},r_{t},s_{t+1},a_{t+1},r_{t+1},....,s_{T},a_{T},r_{T}\}$ under the policy $\pi$. For this trajectory $\tau$, the discounted cumulative rewards can be calculated by
\begin{equation}
	G_{t}(\tau) =r_{t}+\gamma r_{t+1}+\cdots+\gamma^{T-t}r_{T} =\sum_{i=t}^{T}\gamma^{i-t}r_{i}.
\end{equation}

Guided by a policy, multiple trajectories might be generated, and $G_{t}(\tau)$ follows a random distribution. The expected reward is used to evaluate the value of taking $a_{t}$ at $s_{t}$, and the state-action value function is defined as
\begin{equation}
	\resizebox{0.88\hsize}{!}{$
		\begin{aligned}
			Q_{\pi}(s_{t},a_{t}) & = \mathbb{E}_{\pi}[G_{t}\mid S = s_{t},A = a_{t}]  \\
			& = r_{t}+\gamma\mathbb{E}_{\pi}[Q_{\pi}(s_{t+1},a_{t+1})\, \mid S=s_{t},A=a_{t}]. 
		\end{aligned}
		$}
\end{equation}

The goal of DRL is to learn the optimal policy $\pi^{*}$ for maximizing the discounted cumulative reward at any initial state. The optimal action at each state can be located by the optimal state-action value function, which is defined as
\begin{equation}
	Q^{*}(s_{t},a_{t})=\mathrm{max}_{\pi}\,Q_{\pi}(s_{t},a_{t}).
\end{equation}

The value-based DRL (e.g., DQN) uses Deep Neural Networks (DNNs) to approximate $Q^{*}(s_{t},a_{t}; \theta_{Q})$, where $\theta_{Q}$ indicates DNN parameters, learning a deterministic policy by selecting the action with maximum Q-value. However, the huge action space in multi-edge IoV systems seriously affects learning efficiency. Meanwhile, the value-based DRL updates the target network by bootstrapping the Q-network, which is a biased estimation of true action values and may fall into the local optimum. In contrast, the policy-based DRL can better handle the huge action space, which selects actions based on probability distributions. Unless the policy tends to be deterministic, the probability of selecting good actions might be small, causing a high variance in estimating the policy gradient and an unstable training process.

\begin{figure}[!ht]
	\centering
	\begin{center}
		\includegraphics*[width=0.90\linewidth]{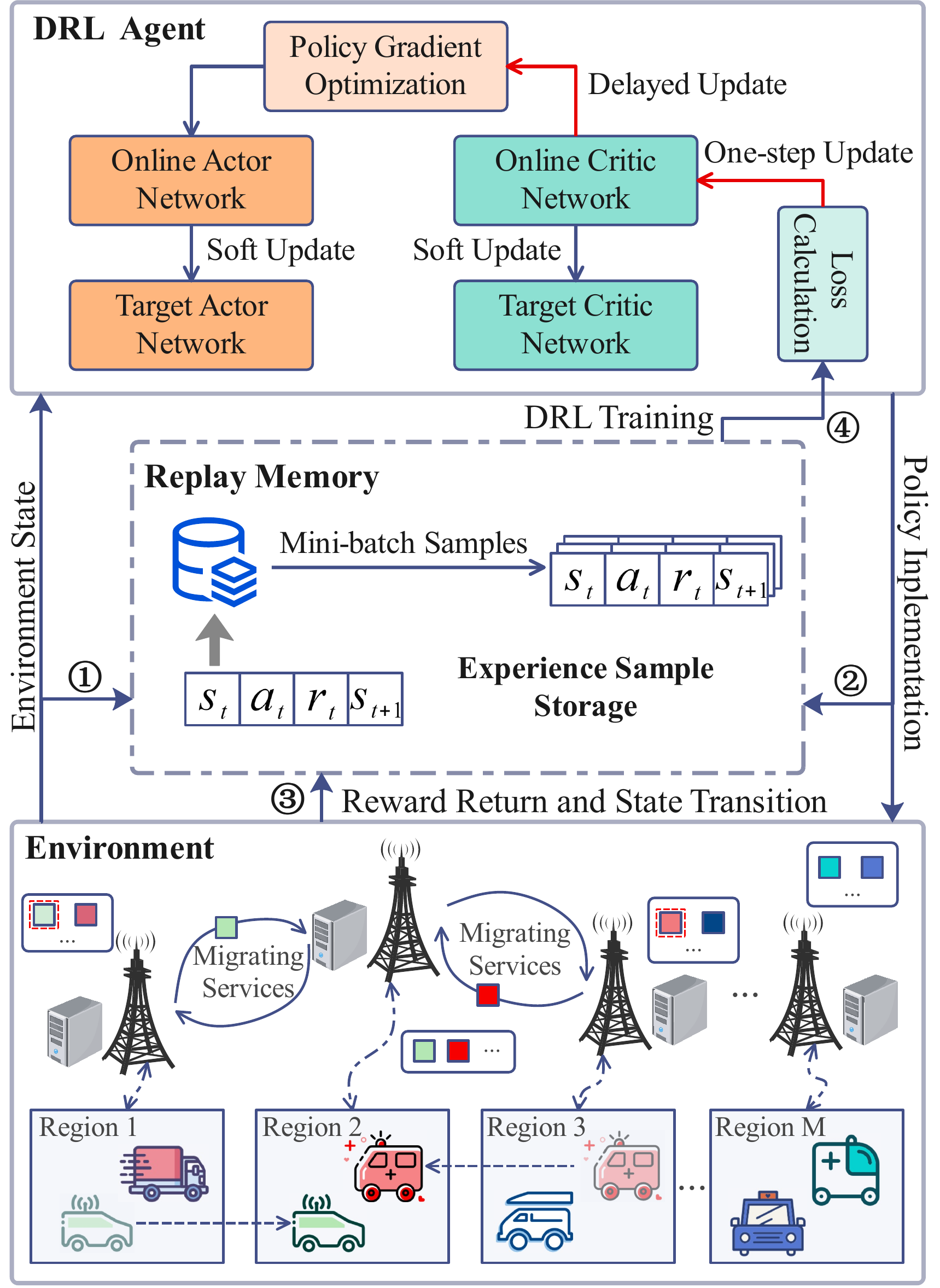}
		  \vspace{-0.2cm}
		\caption{The proposed improved DRL for service migration.}
            \vspace{-0.5cm}
		\label{SRCL_model}
	\end{center}
\end{figure}

To address these issues, we propose an improved actor-critic-based DRL with the asynchronous update to explore the optimal service migration in dynamic multi-edge IoV systems. As illustrated in Fig. \ref{SRCL_model}, during the optimization process of service migration, the DRL agent selects the action $a_{t}$ at the state $s_{t}$ according to the policy $\pi$, and the environment feedbacks the instant reward $r_{t}$ and transits to the next state $s_{t+1}$, which can be expressed as an MDP. The DRL agent then extracts mini-batch samples from the replay memory for network training. Specifically, the state space, action space, and reward function are defined as follows.

\textbf{State space:} At the time slot $t$, the state consists of the information about IVs (including the locations of IVs, the data amount of tasks, the computational density of tasks, and the data amount of service data) and the service migration decisions at time slot $t$-1. Therefore, the state is defined as
\begin{equation}
	\resizebox{0.89\hsize}{!}{$s_{t}=\left\{Loc_{u,t},D_{u,t},C_{u,t},S_{u,t},x_{u,t-1} \mid {IV_{u}\in\mathcal{IV}}\right\}.$}
\end{equation}

\textbf{Action space:} At the time slot $t$, the DRL agent takes an action of service migration $a_{t}$ at $s_{t}$. For $\textstyle IV_{u}$, its service instance can be migrated to another edge node. Therefore, the action is defined as
\begin{equation}
	a_{t}=\left\{x_{u,t} \mid {IV_{u}\in{\mathcal{IV}}}\right\}.
\end{equation}

\textbf{Reward function:} The reward is negatively correlated with long-term system delays including migration, communication, and computation delays, and it is defined as
\begin{equation}
	r_{t}=-\sum_{IV_{u} \in \mathcal{IV}}\left(MT_{u,t}+HT_{u,t}+CT_{u,t}\right).
\end{equation}

The proposed method adopts an actor-critic architecture and deep deterministic policy gradient to train and optimize service migration policies in complex and dynamic multi-edge IoV systems. Specifically, the actor generates actions of service migration, while the critic evaluates the Q-values of actions. The critic with one-step update enables fast convergence and accurately evaluates the actions, which effectively guides the actor's update and significantly reduces the errors when estimating the policy gradient.

\begin{algorithm}
	\caption{Improved DRL for service migration}
	\label{alg:ddpg}

    \KwIn{Actor network $\mu$ , critic network $\psi$, target actor network $\hat{\mu}$, target critic network $\hat{\psi}$, and the multi-edge IoV system }

    \KwOut{Well-trained actor network $\mu$}
	
	\textbf{Initialize:} $\mu$ and $\psi$ with $\theta_{\mu}$ and  $\theta_{\psi}$
	
	\textbf{Initialize:} $\hat{\mu}$ and $\hat{\psi}$ with $\theta_{\hat{\mu}}\leftarrow\theta_{\mu}$ and $\theta_{\hat{\psi}}\leftarrow\theta_{\psi}$
	
	\textbf{Initialize:} Replay memory $X$, training episodes $E$, and maximum time slot per episode $T$  
	
	\For{{\rm $episode = 1, 2, ..., E$}}
	{
		Get initial state: $s_{0}=env.reset()$;
		
		Create service instances on edge nodes;
		
		\For{{\rm $t = 1, 2, ..., T$}}{
			Update IVs' locations and generate service migration decision $a_{t}$: $a_{t}=\mu(s_{t}\mid\theta_{\mu})+ \xi\;$;
			
			Execute service migration decision $a_{t}$;
			
			\For{{\rm $each$ $edge$ $node$ $ f_{m} \in \mathcal{F}$ $\textbf{in}$ $\textbf{parallel}$}}{
				Derive the optimal resource allocation via convex optimization theory (Section 4.2);
			}
			
			Calculate reward and perform state transition: $r_{t}, s_{t+1} = env.step(a_{t})$;
			
			Store training sample: $X.push(s_{t}, a_{t}, r_{t}, s_{t+1})$;
			
			Draw $N$ samples from $X$: $N^{*}(s_{t}, a_{t}, r_{t}, s_{t+1}) = X.sample(N)$;
			
			Calculate cumulative expected rewards: $y_{t}=r_{t}+\gamma\hat{\psi}(s_{t+1},\hat{\mu}(s_{t+1}\mid\theta_{\hat{\mu}})\mid\theta_{\hat{\psi}})\;$;
			
			Calculate loss function and update critic: $Loss_{\psi} = \frac{1}{N} {\textstyle \sum_{1}^{N}} (y_{t} - \psi(s_{t}, a_{t} \mid \theta_{\psi}))^{2}$;
			
			\If{{\rm $t\%\lambda = 0 $}}{
				Update actor using gradient ascent: $\nabla_{\theta_{\mu}}J=\frac{1}{N}\sum_{1}^{N}\nabla_{{a_{t}}}\psi(s_{t},a_{t}\mid\theta_{\psi})$ $\nabla_{\theta_{\mu}}\mu(s_{t}\mid\theta_{\mu})\mathrm{~;~}$
				
				Soft update target network parameters: $\theta_{\hat{\mu}}\leftarrow\omega\theta_{\mu}+(1-\omega)\theta_{\hat{\mu}},$ $\theta_{\hat{\psi}}\leftarrow\omega\theta_{\psi}+(1-\omega)\theta_{\hat{\psi}}$;
			}
		}
	}
\end{algorithm}

The key steps of the proposed method are given in Algorithm 1. 
First, we initialize the actor network $\mu$, critic network $\psi$, target actor network $\hat{\mu}$, target critic network $\hat{\psi}$, replay memory $X$, number of training episodes $E$, and the maximum time slots per episode $T$ (Lines 1$\sim$3). 
For each training episode, IVs create service instances on their nearest edge nodes (Lines 5$\sim$6). 
For the time slot $t$, the system state $s_{t}$ is input into the actor $\mu$, which generates and executes an action of service migration $a_{t}$. Specifically, the Gaussian noise $\xi$ is introduced to enhance the exploration capability of the policy in the initial training phase. Meanwhile, to ensure the stable convergence of the policy in the later training phase, the variance of $\xi$ gradually declines with the increase of time slots, thereby achieving a balance between exploration and convergence. (Lines 8$\sim$9). 
Next, we derive the optimal resource allocation for the service instances on each edge node via convex optimization theory (Lines 10$\sim$12), where the details are given in Section 4.2. 
Then, we calculate the reward $r_{t}$ and the system state transits to $s_{t+1}$ (Line 13). Next, the training sample $(s_{t}, a_{t}, r_{t}, s_{t+1})$ is stored into the replay memory $X$, and we randomly draw $N$ samples from $X$ to train network parameters (Lines 14$\sim$15). 
Notably, the correlation of training samples is broken by using the replay memory, which alleviates the instability that occurs in the training process. Then, we calculate the cumulative expected discount rewards by combining the reward $r_{t}$ with the target critic $\hat{\psi}$, and we adopt the Adam optimizer to minimize the loss of the critic $\psi$ (Lines 16$\sim$17), enabling the updated critic can better fit in $Q^{*}(s_{t},a_{t})$. 
However, the critic is prone to instability, which also leads to fluctuations in the update of the actor.  
To solve this issue, we design a delayed update mechanism, where the actor is updated only after the critic has been updated $\lambda$ times (Lines 18$\sim$21). Specifically, the actor is used to fit a high-dimensional mapping from $s_{t}$ to $a_{t}$, whose objective is defined as

\begin{equation}
	J(\theta_{\mu}) = \mathbb{E}_{\theta_{\mu}}[\psi(s_{t},\mu(s_{t}\mid\theta_{\mu})\mid\theta_{\psi})].
\end{equation}

For a given state $s_{t}$, the actor is updated by adjusting $\theta_{\mu}$, and thus its output $\mu(s_{t}\mid\theta_{\mu})$ can be updated in an upward direction according to $\psi(s_{t},\mu(s_{t}\mid\theta_{\mu})\mid\theta_{\psi})$ calculated by the critic. 
Specifically, the gradient ascent is used to update the actor, and the network parameters of the target actor and the target critic are updated via the soft update (Lines 19$\sim$20). 
In addition, we incorporate gradient clipping during the network update to enhance model stability.

\subsection{Resource Allocation with Convex Optimization}
As aforementioned, \textit{P}3 is to address the problem of resource allocation for all MEC servers, where the resource allocation processes of different MEC servers are independent and parallel. In this regard, we further separate the problem of resource allocation for each MEC server from \textit{P}3, which can be formulated as
\begin{equation}
	\label{p4}
	\begin{aligned}
		\textit{P}4: \min_{e_{u,t}} &  \sum_{IV_{u} \in \mathcal{IV}_{m}} CT_{u,t} = \min_{e_{u,t}} &  \sum_{IV_{u} \in \mathcal{IV}_{m}} {\frac{K_{u,t}}{e_{u,t}F}} \\
		\text { s.t. } & \quad C 2- C3 .
	\end{aligned}
\end{equation}

Therefore, the solution of \textit{P}3 relies on addressing \textit{P}4.
To optimize \textit{P}4, we design a new method based on convex optimization theory. 
First, we prove that \textit{P}4 is a convex programming problem with the constraints of \textit{P}4 and Hessian matrix. 
Next, guided by convex optimization theory, we define the generalized Lagrange function and solve the KKT conditions for \textit{P}4. 
Finally, we theoretically derive the optimal resource allocation for each MEC server under given decisions of service migration. 

\textbf{Lemma 2.} \textit{P}4 \textit{is a convex programming problem}.

\textbf{Proof.} This lemma can be proved if \textit{P}4 and its constraints are both convex functions. It can be observed that \textit{C}2 and \textit{C}3 are linearly constrained, reflecting the convexity. Thus, we only need to prove that \textit{P}4 is also a convex function.

To clarify the derivation process, we reassign the index of $\mathcal{IV}_{m}$ into $\mathcal{Z} = \{IV_{1}, IV_{2}, ..., IV_{z}, ..., IV_{Z}\}$ and rewrite \textit{C}2 accordingly. Therefore, Eq. (\ref{p4}) can be redefined as
\begin{equation}\label{re_p3}
	\begin{aligned}
		\mathcal{Y} [e_{1,t},& e_{2,t}, ..., e_{Z,t}] = \min _{e_{z,t}}  \sum_{IV_{z} \in \mathcal{Z}} \frac{K_{z,t}}{e_{z,t}F} \\
		\text { s.t. } \quad C 3 & : \sum_{IV_{z} \in \mathcal{Z}} e_{z,t}-1\leq0, t \in T, IV_{z} \in \mathcal{Z}, \\
		C 4 & : -e_{z,t} \leq 0, t \in T, IV_{z} \in \mathcal{Z}, \\
		C 5 & : e_{z,t} - 1 \leq 0, t \in T, IV_{z} \in \mathcal{Z} ,
	\end{aligned}
\end{equation}
where \textit{C}2 is converted into \textit{C}4 and \textit{C}5, which indicate the upper and lower bounds of the proportion of allocated resources, respectively.

The Hessian matrix is a square matrix that is established by the second-order partial derivatives of a multivariate function, which reflects the local curvature of a function. The Hessian matrix of Eq. (\ref{re_p3}) is defined as
\begin{equation}
	\resizebox{0.89\hsize}{!}{$
		H=\left[\begin{array}{cccc}
			\frac{\partial^{2} \mathcal{Y}}{\partial\left(e_{1,t}\right)^{2}} & \frac{\partial^{2} \mathcal{Y}}{\partial e_{1,t} \partial e_{2,t}} & \cdots & \frac{\partial^{2} \mathcal{Y}}{\partial e_{1,t} \partial e_{Z,t}} \\
			\frac{\partial^{2} \mathcal{Y}}{\partial e_{2,t} \partial e_{1,t}} & \frac{\partial^{2} \mathcal{Y}}{\partial\left(e_{2,t}\right)^{2}} & \cdots & \frac{\partial^{2} \mathcal{Y}}{\partial e_{2,t} \partial e_{Z,t}} \\
			\vdots & \vdots & \ddots & \vdots \\
			\frac{\partial^{2} \mathcal{Y}}{\partial e_{Z,t} \partial e_{1,t}} & \frac{\partial^{2} \mathcal{Y}}{\partial e_{Z,t} \partial e_{2,t}} & \cdots & \frac{\partial^{2} \mathcal{Y}}{\partial\left(e_{Z,t}\right)^{2}}
		\end{array}\right],
		$}
\end{equation}
where
\begin{equation}
	\frac{\partial^{2} \mathcal{Y}}{\partial e_{z_{1},t} \partial e_{z_{2},t}}=\left\{\begin{array}{ll}
		0, & z_{1} \neq z_{2} \\
		\frac{2K_{z_{1},t}}{(e_{z_{1},t})^{3}F}, & z_{1} = z_{2}
	\end{array},\right.
\end{equation}
and $\forall IV_{z_{1}},IV_{z_{2}}\in\mathcal{Z}$, $K_{(\cdot), t} \geq 0$ and $e_{(\cdot),t} \geq 0$.

$F$ is a non-zero real number. Moreover, the values on the diagonal of $H$ are all positive, and thus it is a symmetric positive definite matrix. Based on the above analysis and convex optimization theory \cite{boyd2004convex26}, we prove that \textit{P}4 is a convex programming problem.

Following the optimality theorem of convex programming, any feasible KKT points can be the global optimum. Further, the generalized Lagrange function is defined as
\begin{equation}
	\begin{aligned}
		L =& \sum_{IV_{z} \in \mathcal{Z}}\frac{K_{z,t}}{e_{z,t}F} + \beta(\sum_{IV_{z} \in \mathcal{Z}} e_{z,t}-1) \\ &+ \sum_{IV_{z} \in \mathcal{Z}} \eta_{z}(-e_{z,t})
		 + \sum_{IV_{z} \in \mathcal{Z}}\zeta_{z}(e_{z,t}-1),
	\end{aligned}
\end{equation}
where $\beta,\eta_{z},\zeta_{z},IV_{z}\in\mathcal{Z}$ are the Lagrange multipliers associated with \textit{C}3, \textit{C}4, and \textit{C}5, respectively.

Therefore, the KKT conditions of the redefined \textit{P}4 in Eq. (\ref{re_p3}) can be described as
\begin{equation}
	\begin{cases}
		-\frac{K_{z,t}}{(e_{z,t})^{2}F}+\beta-\eta_{z}+\zeta_{z} = 0, 
		\\
		\beta(\sum_{IV_{z} \in \mathcal{Z}} e_{z,t}-1)=0, \sum_{IV_{z} \in \mathcal{Z}} e_{z,t}-1 \leq 0, 
		\\
		\eta_{z}(-e_{z,t})=0, -e_{z,t}\leq 0, 
		\\
		\zeta_{z}(e_{z,t}-1)=0, e_{z,t}-1\leq 0, 
		\\
		\beta \geq 0, \eta_{z} \geq 0, \zeta_{z} \geq 0. 
		\\
	\end{cases}
\end{equation}

According to the above KKT conditions, the optimal resource allocation can be derived by

\begin{equation}
	\label{res_result}
	e_{z,t}=\frac{\sqrt{K_{z,t}}}{\sum_{IV_{z} \in \mathcal{Z}}\sqrt{K_{z,t}}}, IV_{z} \in \mathcal{Z}.
\end{equation}

By using the proposed method, $\forall t\in T$, each MEC server can obtain optimal resource allocation for service instances (Lines 10$\sim$12 in Algorithm 1), supporting the DRL agent learning better migration policies.

\subsection{Complexity Analysis of \textit{SR-CL}}
For each training step of service migration in Algorithm 1, $N$ samples are extracted from $X$ to train network parameters, involving forward computation and backward propagation. Referring to \cite{du2023maddpg}, the time complexity of a DRL-based algorithm mainly depends on the network structures. Specifically, the actor takes the state $s_{t}$ as input and outputs the action $a_{t}$, while the critic takes the concatenated vector of $s_{t}$ and $a_{t}$ as input and outputs the Q-values. We define the dimensions of $s_{t}$, $a_{t}$, and the two hidden layers of neural networks as $\mathcal{D}$, $\mathcal{W}$, $\mathcal{H}_1$, and $\mathcal{H}_2$, respectively. Therefore, the time complexity of training the actor using a sample is
\begin{equation}
	O_{actor} = O\left ( \mathcal{D}\mathcal{H}_1 + \mathcal{H}_1\mathcal{H}_2 +\mathcal{H}_2\mathcal{W}\right ).
\end{equation}

Similarly, the time complexity of training the critic is
\begin{equation}
	O_{critic} = O\left ( \left ( \mathcal{D} +\mathcal{W} \right )\mathcal{H}_1 + \mathcal{H}_1\mathcal{H}_2 +\mathcal{H}_2\right ).
\end{equation}

In Algorithm 1, the critic is updated every time step, while the actor is updated only after the critic has been updated $\lambda$ times, where their update frequencies are denoted as $TE$ and $TE/\lambda$, respectively. Therefore, the time complexity of Algorithm 1 can be calculated by

\vspace{-0.3cm}
\begin{equation}
\begin{aligned}
	O_{agent} = &O\Bigg ( NTE \Bigg ( (\mathcal{D} + \mathcal{W}+\frac{\mathcal{D}}{\lambda}) \mathcal{H}_{1}\\
	&+ (1+\frac{1}{\lambda}) \mathcal{H}_{1}\mathcal{H}_{2}  + (1+\frac{\mathcal{W}}{\lambda}) \mathcal{H}_{2} \Bigg ) \Bigg ).
\end{aligned}
\end{equation}

For the resource allocation that is performed according to Eq. (\ref{res_result}), its time complexity is $O$(1). If both service migration and resource allocation are incorporated into the action space of DRL, the time complexity will become

\vspace{-0.3cm}
\begin{equation}
\begin{aligned}
	O_{agent}^{'} =& O\Bigg ( NTE \Bigg ( (\mathcal{D} + \mathcal{W}+\mathcal{W}_{RA}+\frac{\mathcal{D}}{\lambda}) \mathcal{H}_{1}\\
	&+ (1+\frac{1}{\lambda}) \mathcal{H}_{1}\mathcal{H}_{2}  + (1+\frac{\mathcal{W}+\mathcal{W}_{RA}}{\lambda}) \mathcal{H}_{2} \Bigg ) \Bigg ),
\end{aligned}
\label{tc_RA}
\end{equation}
where $\mathcal{W}_{RA}$ is the dimension of resource allocation.

It is noted that the time complexity mainly depends on the first term in Eq. (\ref{tc_RA}), and this term will grow rapidly if the action and state spaces become huge. Moreover, the decision space will become hybrid, and it is impractical to conduct thorough exploration in a continuous space. The strict constraints of resource allocation make it challenging to find satisfactory results, leading to massive invalid exploration and thus diminishing learning efficiency. 

Therefore, by combining DRL with convex optimization theory for addressing the joint problem of service migration and resource allocation in multi-edge IoV systems, the proposed \textit{SR-CL} not only can efficiently reach the optimal policy but also effectively reduce the exploration complexity.

\subsection{Implementation of \textit{SR-CL} in Real-world Scenarios}

\begin{figure}[!ht]
	\centering
	\begin{center}
		\includegraphics*[width=0.7\linewidth]{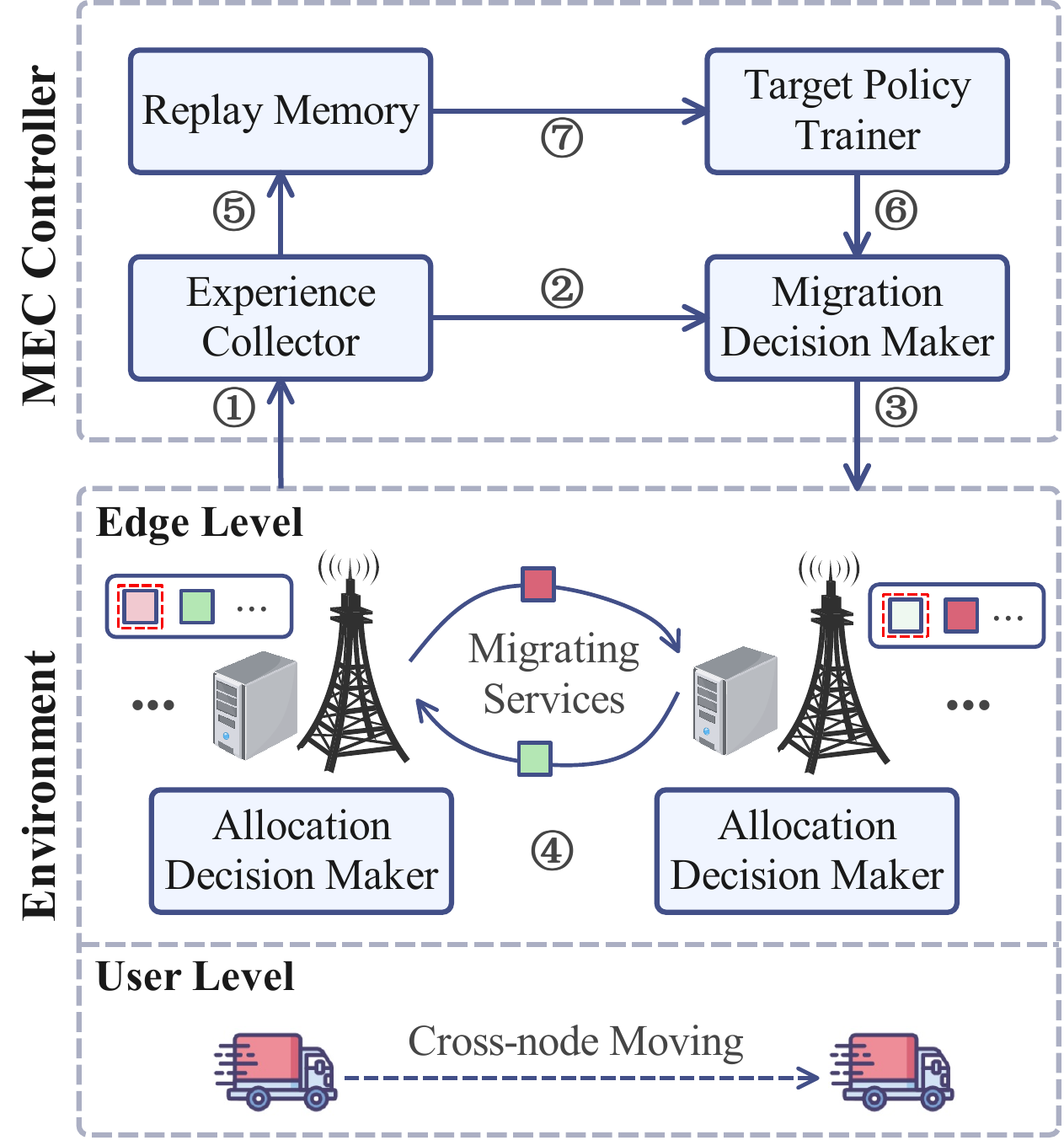}
		\vspace{-0.2cm}
		\caption{Implementation of \textit{SR-CL} in real-world scenarios.}
		\vspace{-0.5cm}
		\label{SRCL_Implementation}
	\end{center}
\end{figure}

As illustrated in Fig. \ref{SRCL_Implementation}, the proposed \textit{SR-CL} contains five main components including experience collector, replay memory, migration decision maker, target policy trainer, and allocation decision maker. The allocation decision maker is deployed on each edge node, while the others are deployed in the MEC controller. Specifically, the experience collector receives the state and reward information from the environment (Step 1) and sends the state to the migration decision maker (Step 2). The migration decision maker consists of the online critic and actor, which synchronizes parameters from the target policy trainer (Step 6) and makes migration decisions based on the state (Step 3). After executing the migration decisions, both tasks and service instances are transmitted to target edge nodes. Meanwhile, the allocation decision makers on each edge node perform optimization in parallel to obtain the optimal resource allocation (Step 4). Next, the experience collector sends the collected experience samples to the replay memory (Step 5), and then the target policy trainer periodically extracts samples from the replay memory for updating policies (Step 7). Therefore, when the network topology or traffic pattern changes, the proposed \textit{SR-CL} can respond to the changes in time and maintain high generality.

\section{Performance Evaluation}\label{sec:experiments}

In this section, we evaluate and analyze the proposed \textit{SR-CL} through extensive simulation and testbed experiments.
\subsection{Experiment Setup}

\begin{figure}[!ht]
	\centering
	\begin{center}
		\includegraphics*[width=0.78\linewidth]{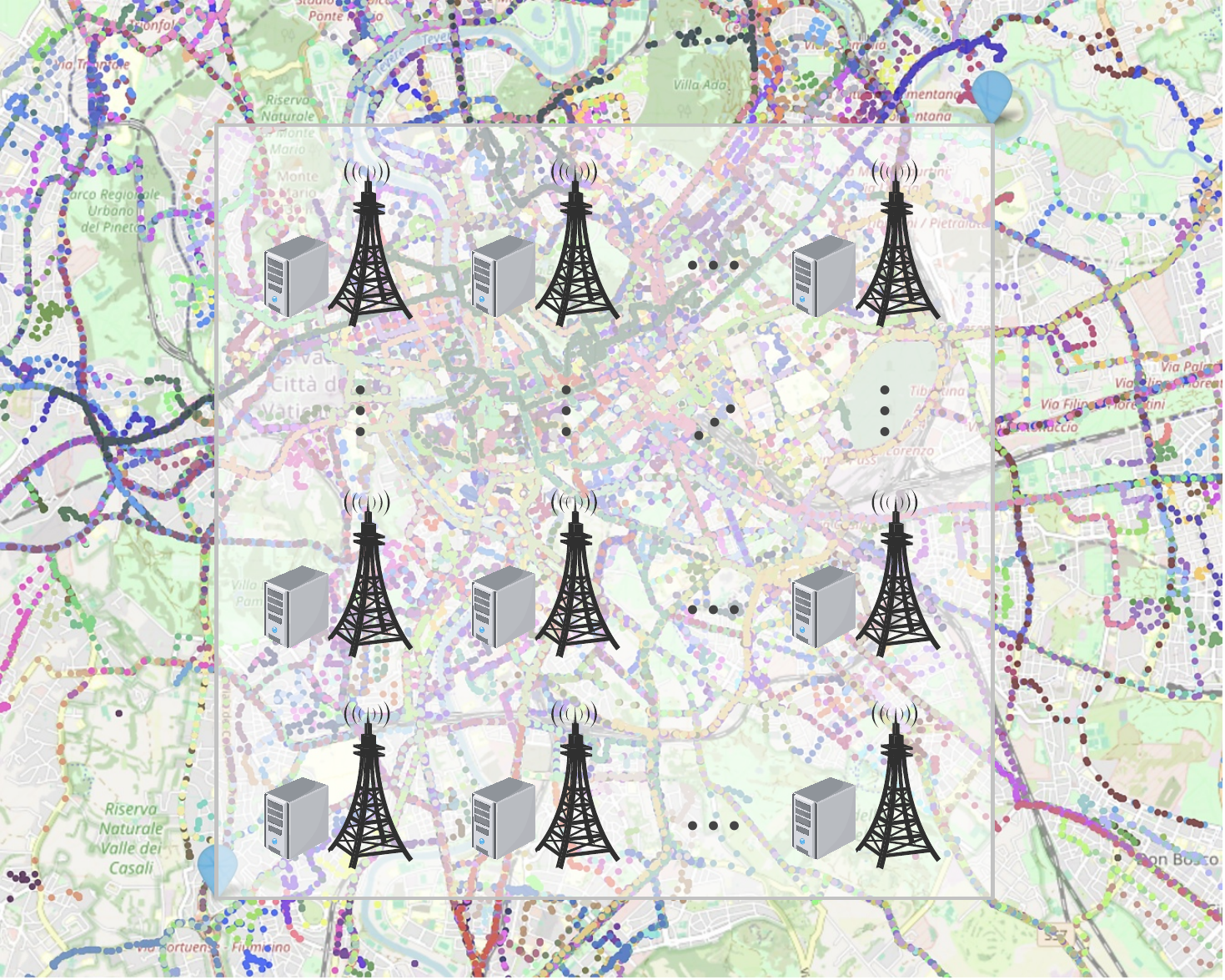}
            \vspace{-0.2cm}
		\caption{Real-world vehicle trajectories and BS distribution.}
            \vspace{-0.5cm}
		\label{datasets}
	\end{center}
\end{figure}

\textbf{Datasets.} 
We evaluate the performance of the proposed \textit{SR-CL} using the real-world datasets of vehicle trajectories in Rome City \cite{Bracciale2022}. The datasets contain the driving data from 320 taxis over 30 days, spanning from February 1st, 2014, to March 2nd, 2014. Each record includes the unique taxi ID, timestamp, and trajectory coordinates. As illustrated in Fig. \ref{datasets}, the scatter points with different colors indicate the trajectories of different vehicles in cities. By tracking the vehicle trajectories, we can obtain the sources and locations of different vehicles’ tasks and then simulate dynamic and varying vehicle mobility patterns.


\begin{figure*}
	\centering
	\begin{minipage}[t]{0.335\linewidth}
		\centering
		\includegraphics[width=2.4in]{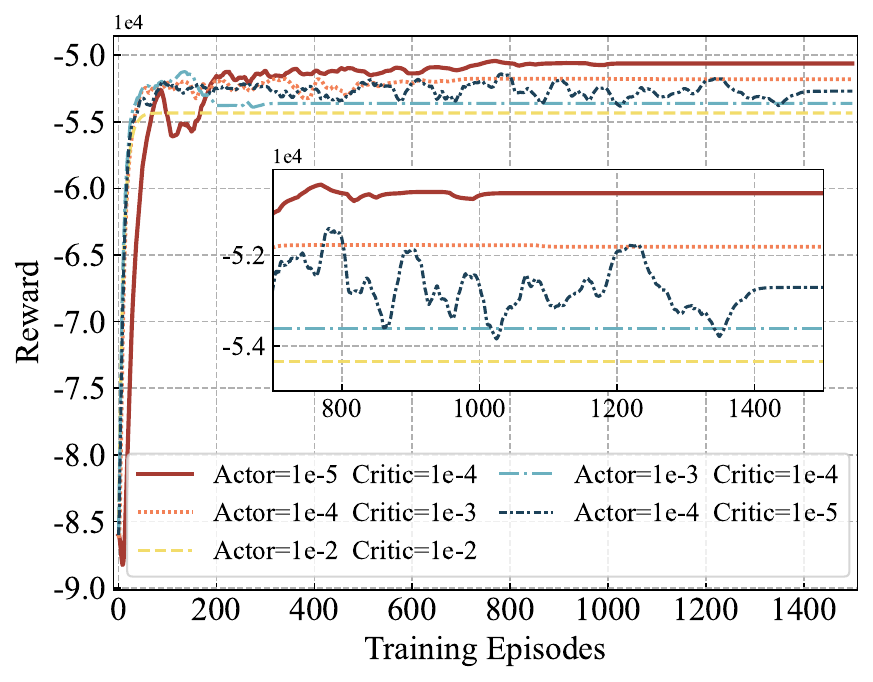}
            \vspace{-0.7cm}
		\caption{Convergence vs. learning rate.}
		\label{Learning_rate}
	\end{minipage}%
	\begin{minipage}[t]{0.33\linewidth}
		\centering
		\includegraphics[width=2.4in]{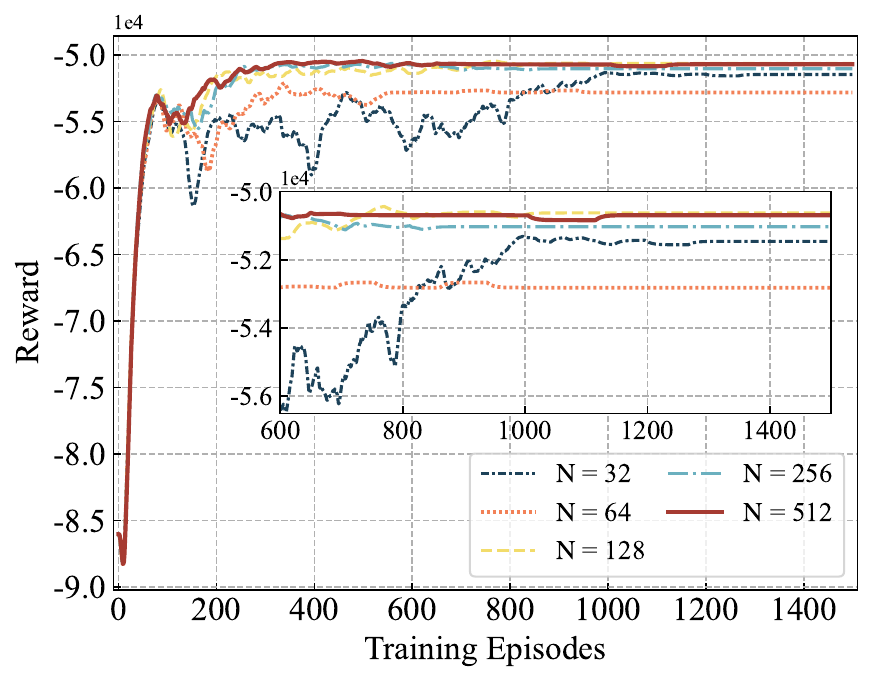}
            \vspace{-0.7cm}
		\caption{Convergence vs. batch size.}
		\label{Batch_size}
	\end{minipage}
	\begin{minipage}[t]{0.33\linewidth}
		\centering
		\includegraphics[width=2.4in]{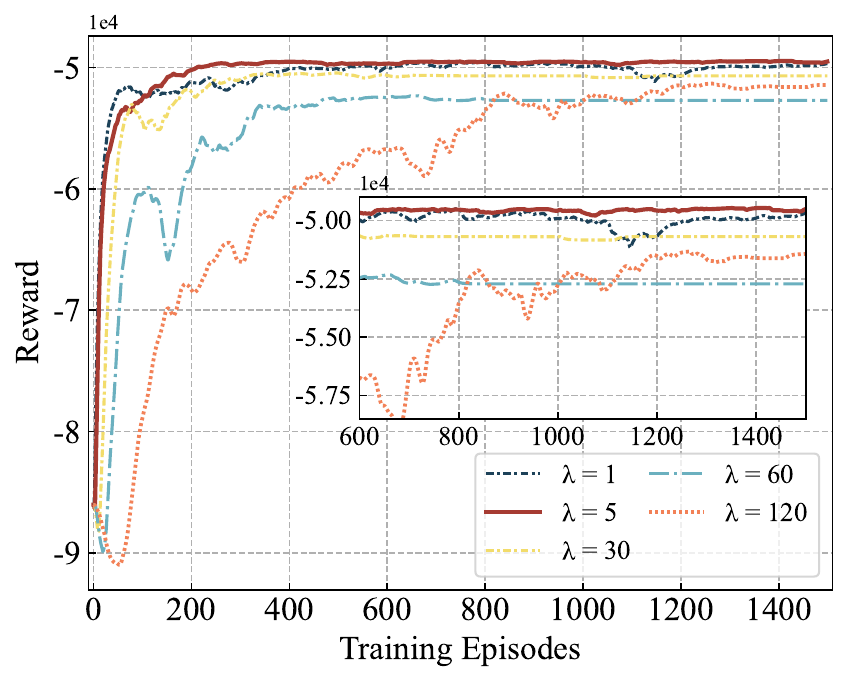}
            \vspace{-0.7cm}
		\caption{Convergence vs. update frequency.}
		\label{Update_frequency}
	\end{minipage}
\end{figure*}
\textbf{Simulation Settings.}
The simulation experiments are conducted on a workstation with one 8-core Intel(R) Xeon(R) Silver 4208 CPU @3.2GHz and 2 NVIDIA GeForce RTX 3090 GPUs with 32GB RAM. As shown in Fig. \ref{datasets}, we consider the region enclosed by the coordinate pairs [41.856, 12.442] and [41.928, 12.5387] of the city center as our experiment scenario, which encompasses the complex mobility patterns of numerous vehicles. To facilitate our analysis, we scale down this region to 16 $\text{km}^{2}$ area served by 16 edge nodes. The communication coverage of each edge node is 1 $\text{km}^{2}$. The bandwidth of each BS is 20 MHz, and the computational capability of each MEC server is 60 GHz. For the default experiment settings, we consider a high-connectivity network topology, where each node can communicate with its adjacent edge nodes. The traffic load distribution among edge nodes dynamically changes with migration decisions. At the initial, there are 100 IVs connected to their nearest edge nodes, and the MEC servers create service instances for IVs. An episode consists of 240 time slots, and IVs continuously send tasks to their service instances at fixed intervals. Based on Python 3.8 and Pytorch, we build and train the neural networks in \textit{SR-CL}, which owns two hidden layers with 512 and 256 neurons, respectively. Moreover, the batch size $N$ is 512, the delayed update parameter $\lambda$ is 5, the soft update parameter $\omega$ is 1e-2, the learning rate of the actor and critic are 1e-5 and 1e-4, the gradient clipping value is 2.0, the Gaussian noise is $\xi \sim \mathcal{N}(0,0.15^{2})$, the size of replay memory is 1e4, and the discount factor $\gamma$ is 0.95. After the \textit{SR-CL} completes training, it can be applied to various scenarios. Other main parameters are listed in Table \ref{table_parameter}.


\newcommand{\upcite}[1]{\textsuperscript{\cite{#1}}}
\begin{table}[!ht] 
	\renewcommand{\arraystretch}{1.3}
	\caption{Settings of simulation parameters}
	\label{table_parameter}
	\centering
	\begin{tabular}{|c||c|}
		\hline
		{\textbf{Parameter}} &  
		\tabincell{c} 
		{\textbf{Default value}} \\
		\hline\hline
		Transmission power, $P_{u}$ & $\mathcal{U}(0.4, 0.6)$ W  \upcite{chai2021dynamic28}\\
		
		\hline
		Data amount of each task, $D_{u,t}$ & $\mathcal{U}(0.5, 1.5)$ MB \upcite{chai2021dynamic28} \\ 
		
		\hline
		Computational density, $C_{u,t}$ & $\mathcal{U}(200, 1000)$ cycles/bit \upcite{wang2021computation29} \\ 
		
		\hline
		Data amount of service data, $S_{u,t}$ & $\mathcal{U}(0.5, 50)$ MB \upcite{zhang2022deep18, wang2022online27} \\ 
		\hline
		Gaussian noise, $\sigma^{2}$ & $10^{-13}$ W \upcite{wang2021computation29}\\
		
		\hline
		Channel gain per unit distance, $\alpha$  & $10^{-5}$ \upcite{wang2021computation29} \\
		\hline
		Transmission delay coefficient, $\mu^{h}$ & $0.3$ s\\
		
		\hline
		Migration delay coefficient, $\mu^{y}$ & $1.5$ s\\	
		
		\hline
		Network bandwidth, $\chi$ & $500$ Mpbs \upcite{wang2022online27} \\
		
		\hline
	\end{tabular}
\end{table}

\textbf{Comparison Methods.}
We compare the \textit{SR-CL} with the following benchmark methods to verify its superiority. 

\begin{itemize}
	\item \textit{Always Migrate (AM)} \cite{ouyang2018follow8}: Service instances are always migrated to the nearest edge nodes along with IVs.
	\item \textit{Never Migrate (NM)} \cite{wang2022online27}: Service instances are always located at the initial edge nodes.
	\item \textit{Genetic Algorithm (GA)} \cite{maia2021improved}: The reward is defined as the fitness value, and a heuristic evolutionary algorithm is used to search for the optimal migration individual.
	\item \textit{Independent Deep Q-Network (IDQN)}: As an extension of Independent Q-Learning (IQL) \cite{yao2023cooperative}, the \textit{IDQN} utilizes DNNs to approximate the Q-function. In the experiments, DRL agents are deployed on vehicles that make migration decisions based on local observations, which learn deterministic policies and select the actions with the maximum Q-value. There is no communication process among different DRL agents.
	\item \textit{Joint optimization of Service migration and Resource allocation (JSR)}: To verify the validity of the proposed problem decoupling, DRL is individually used to optimize the joint problem of service migration and resource allocation. In the experiments, it adopts the same network structure and parameter settings as the \textit{SR-CL}.
	\item \textit{Deep Deterministic Policy Gradient (DDPG)} \cite{lan2020deep15}: As an advanced DRL, it adopts deterministic policy gradient and one-step update to optimize migration policies.
\end{itemize}

In the experiments, all the comparison methods (except the \textit{JSR}) allocate resources to service instances in proportion to task demands. To ensure the fairness of experiments, scenario settings are consistent for all methods.

\subsection{Experiment Results and Analysis}
\subsubsection{Hyperparameter Tunning of SR-CL}

\begin{figure}[!ht]
	\centering
	\begin{center}
		\includegraphics*[width=0.81\linewidth]{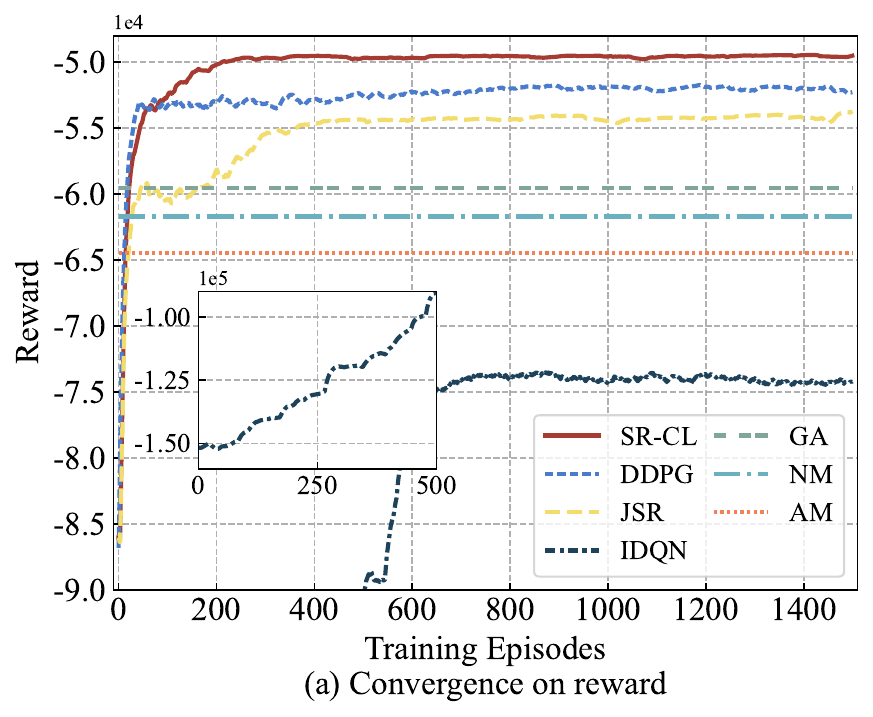}
          \includegraphics*[width=0.80\linewidth]{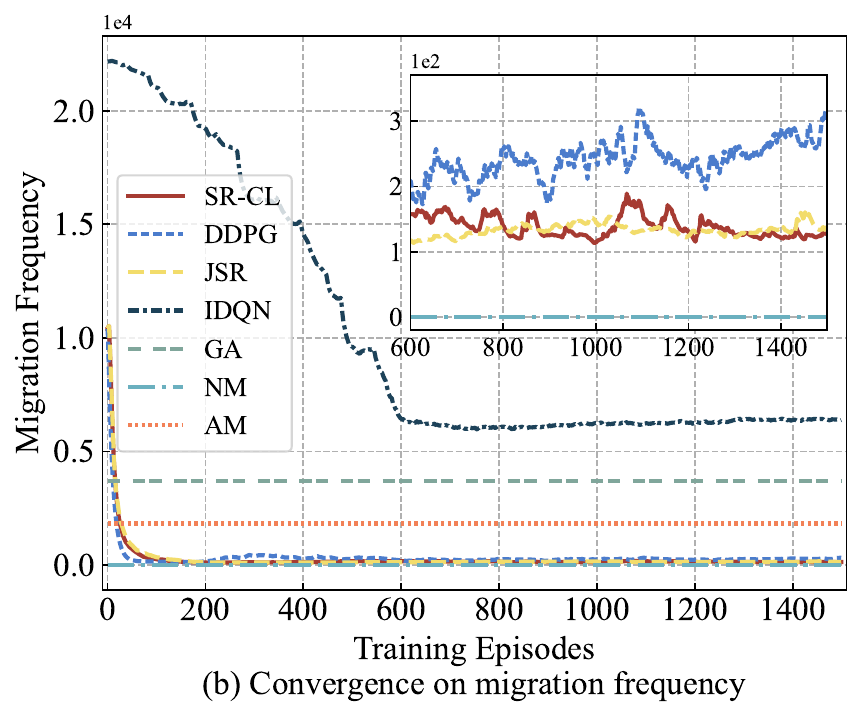}
        \vspace{-0.2cm}
		\caption{Convergence comparison of different methods in terms of reward and migration frequency.}
            \vspace{-0.5cm}
		\label{Convergence}
	\end{center}
\end{figure}

\textbf{Learning Rate.} As depicted in Fig. \ref{Learning_rate}, we evaluate the convergence of the proposed \textit{SR-CL} with different learning rates. It is common practice for the critic to have a higher learning rate than the actor, enabling the critic to approximate the Q-value function more accurately. 
When the critic's learning rate is lower than the actor's, the inaccurate estimation of Q-values leads to the persistent oscillations of the training process. Moreover, an excessively large learning rate may cause premature convergence to the local optimum. By setting the actor's learning rate lower than the critic's, the \textit{SR-CL} achieves stable training and better convergence. Through extensive experiment tests and analysis, we set the critic's and the actor's learning rates in the \textit{SR-CL} to 1e-4 and 1e-5, respectively.

\begin{figure*}
	\centering
	\begin{minipage}[t]{0.33\linewidth}
		\centering
		\includegraphics[width=2.43in]{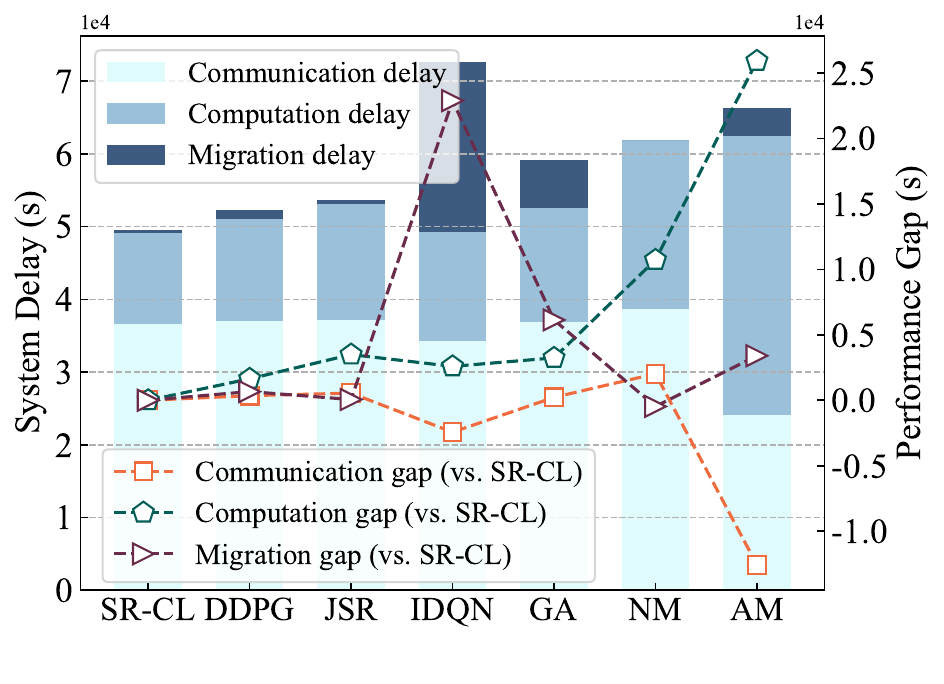}
            \vspace{-0.7cm}
		\caption{Various delays of different methods with default experiment settings.}
		\label{Various_time}
	\end{minipage}
	\begin{minipage}[t]{0.33\linewidth}
		\centering
		\includegraphics[width=2.4in]{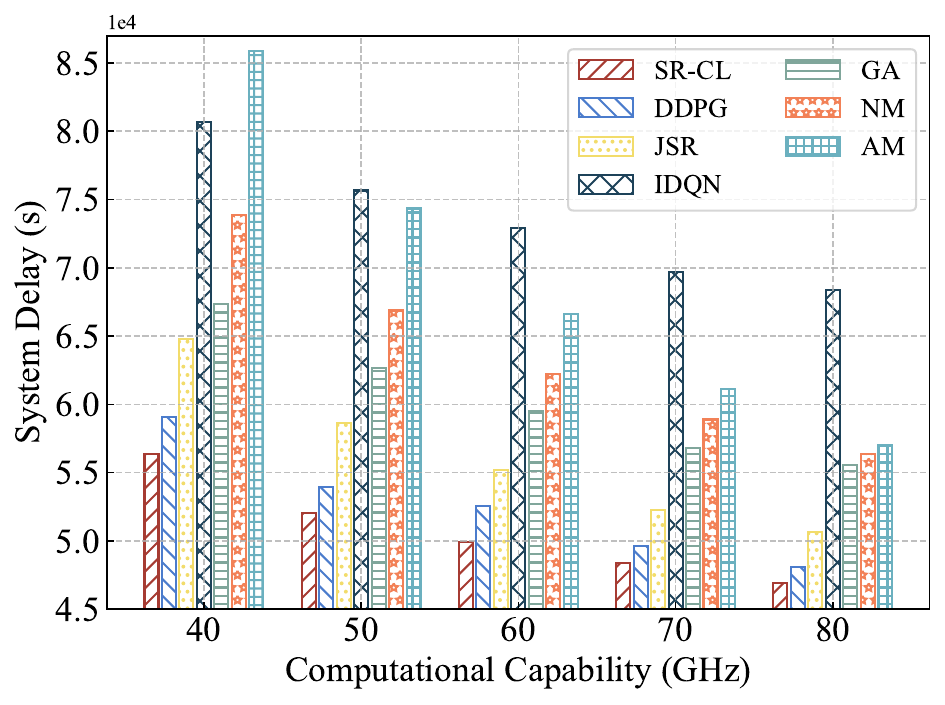}
            \vspace{-0.7cm}
		\caption{Comparison of different methods with various computational capabilities.}
		\label{Computational_capability}
	\end{minipage}
	\begin{minipage}[t]{0.33\linewidth}
		\centering
		\includegraphics[width=2.4in]{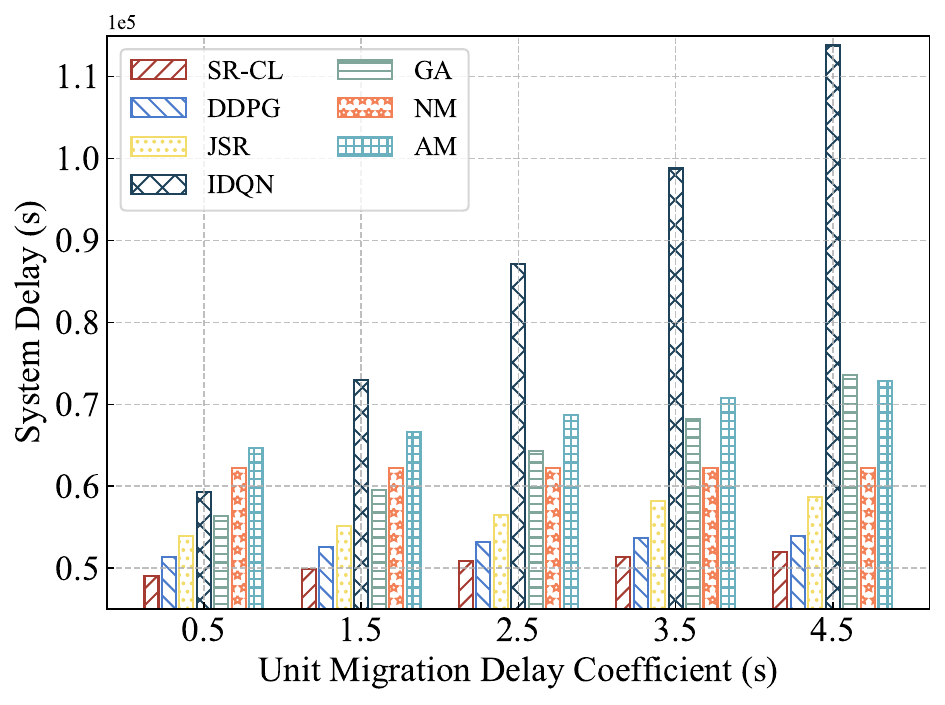}
            \vspace{-0.7cm}
		\caption{Comparison of different methods with various delay coefficients.}
		\label{Migration_hop}
	\end{minipage}
\end{figure*}

\textbf{Batch Size.} As illustrated in Fig. \ref{Batch_size}, we test the effect of different batch sizes on the convergence of the proposed \textit{SR-CL}. The results show that larger batch sizes can offer more sample information and thus improve the model's generalization ability. Moreover, the \textit{SR-CL} with larger batch sizes yields more stable gradient estimation because it can better represent the statistical characteristics of the entire training datasets and reduce gradient variance, thereby stabilizing the training process. Specifically, when the batch size $N$ is set to 32, the training process of the \textit{SR-CL} exhibits significant fluctuations. As the batch size increases, the \textit{SR-CL} achieves more stable convergence performance. Based on the above analysis, we set the batch size $N$ of the \textit{SR-CL} to 512 in the following experiments.

\textbf{Update Frequency.} As shown in Fig. \ref{Update_frequency}, we assess the convergence of the proposed \textit{SR-CL} with different update frequencies. Specifically, when the update frequency $\lambda$ is set too small (e.g., 1), the \textit{SR-CL} implies frequent updates of the actor by using the critic that is not well trained. This leads to inaccurate Q-value estimation, significant update fluctuations, and a potential trap in the local optimum. Conversely, if $\lambda$ is set too large, the training process of the \textit{SR-CL} will become overly sluggish. To strike a good balance between accuracy and efficiency, we set the update frequency $\lambda$ to 5.

\subsubsection{Convergence Comparison of Different Methods}
\textbf{Convergence on Reward.} We compare the convergence of different methods in terms of reward. As shown in Fig. \ref{Convergence}(a), the \textit{AM}, \textit{NM}, and \textit{GA} never involve the training process of neural networks, and thus the reward remains unchanged with the increase of training episodes. The \textit{IDQN} makes migration decisions only based on local observation but ignores other agents in the environment, deviating from the global optimum. The \textit{AM} and \textit{NM} make migration decisions by preset rules, working better than the \textit{IDQN}. The \textit{GA} only considers the instant reward without long-term perspective. In the early stage, the \textit{SR-CL}, \textit{DDPG}, and \textit{JSR} do not show superior performance. As the training progresses, these three methods gradually outperform other methods by continuously exploring and learning. The \textit{JSR} incorporates resource allocation into the action space. However, it is extremely challenging for the \textit{JSR} to fully explore the continuous action space, causing premature convergence to the local optimum. The \textit{DDPG} focuses on optimizing migration decisions and outperforms the \textit{JSR}. Compared to the \textit{DDPG}, the \textit{SR-CL} integrates an improved DRL with convex optimization theory, achieving a more stable training process and superior convergence. 

\textbf{Convergence on Migration Frequency.} We compare the convergence of different methods in terms of migration frequency. As illustrated in Fig. \ref{Convergence}(b), the \textit{IDQN} and \textit{NM} exhibit the highest and lowest migration frequency, respectively. Commonly, IVs tend to stay in the communication coverage of a BS within consecutive time slots, and thus the migration frequency of the \textit{AM} is low. The \textit{GA} always makes migration decisions by searching the randomly initialized population, resulting in higher migration frequency than the \textit{AM}. As the training progresses, the \textit{SR-CL}, \textit{DDPG}, and  \textit{JSR} are able to learn better policies. Due to the one-step update mechanism in \textit{DDPG}, the critic generates inaccurate Q-value estimation for actions, which consequently affects the update of the actor in the right direction. As a result, the migration frequency of the \textit{DDPG} remains fluctuating within a certain range. Compared to the \textit{DDPG}, the \textit{SR-CL} achieves better convergence performance and lower migration frequency, validating the superior design of the \textit{SR-CL}.

\subsubsection{Delay Comparison of Different Methods}
\textbf{Various Delays of Different Methods.} As depicted in Fig. \ref{Various_time}, we compare various delays (including communication, computation, migration, and total delays) of the \textit{SR-CL} with other methods. Since the \textit{NM} does not perform service migration, there is no migration delay. However, the \textit{NM} assembles service instances on the initial edge nodes, leading to a significant delay of the backhaul link transmission as IVs move. The \textit{AM} always migrates service instances along with the movement of IVs, which reduces communication delay but causes the overload issue and increases computation delay. The \textit{IDQN} adopts the strategy of individual independent optimization and fails to capture the inherent complexity of the whole system, resulting in excessive migration delay. Therefore, the performance of the \textit{IDQN} is inferior to other methods. Frequent migrations may happen in \textit{GA}, but it is superior to the \textit{NM}, \textit{AM}, and \textit{IDQN} with the guidance of prior knowledge. Compared to the above methods, the \textit{JSR} achieves better performance. However, due to insufficient exploration in continuous space, it cannot obtain the global optimum. In contrast, the \textit{DDPG} and \textit{SR-CL} outperform other methods, where the \textit{SR-CL} enables the optimal resource allocation by using convex optimization and thus exhibits the lowest total delay.

\begin{figure*}
	\centering
	\begin{minipage}[t]{0.33\linewidth}
		\centering
		\includegraphics[width=2.0in]{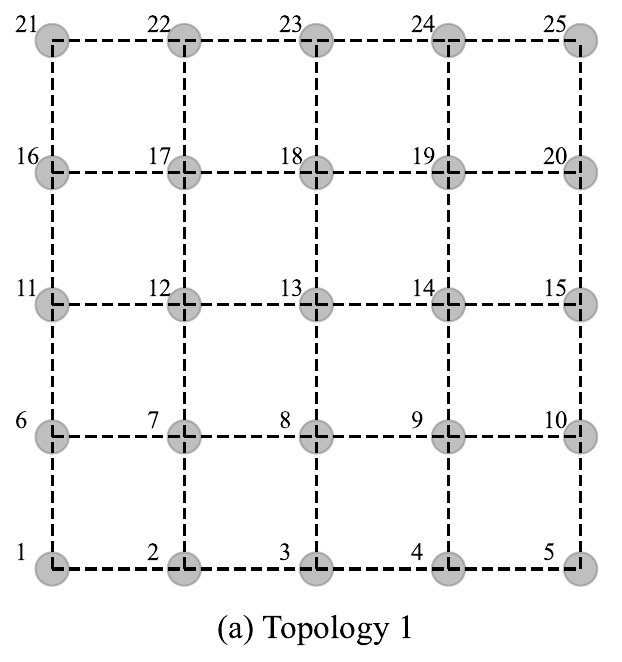}
	\end{minipage}
    \begin{minipage}[t]{0.33\linewidth}
		\centering
		\includegraphics[width=2.0in]{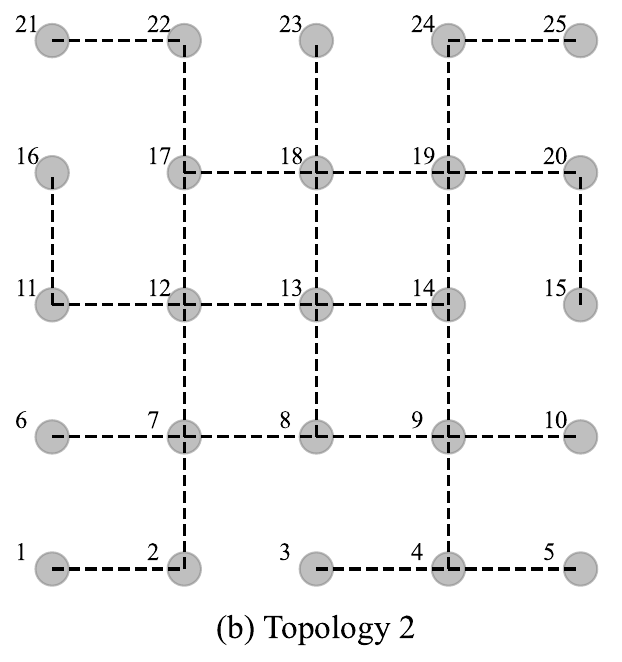}
	\end{minipage}
	\begin{minipage}[t]{0.33\linewidth}
		\centering
		\includegraphics[width=2.0in]{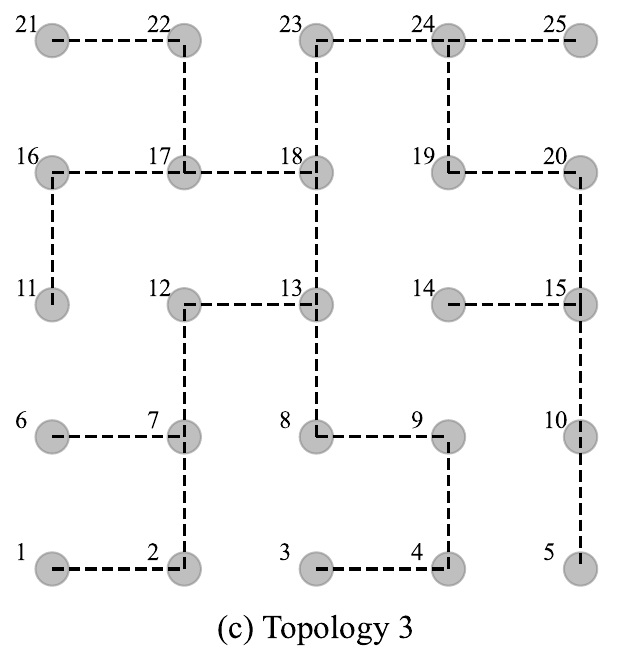}
	\end{minipage}
    \vspace{-0.7cm}
     \caption{Different topologies of a large-scale network with 25 edge nodes.}
	\label{Topologies}
\end{figure*}

\textbf{Total Delay with Various Computational Capabilities.}
As shown in Fig. \ref{Computational_capability}, there is a downward trend in total system delay as the computational capability of each MEC server increases. This is because MEC servers can provide IVs with more available computational resources, thereby reducing the computation delay. For the \textit{AM}, the total system delay primarily depends on computation delay caused by the overload issue (as analyzed in Fig. \ref{Various_time}), and thus the increase of computational capabilities can better alleviate it. However, the high communication delay in \textit{NM} can not be mitigated, limiting the performance improvement. The \textit{GA} and \textit{IDQN} are able to find suitable migration strategies when the computational capability is low. With the increase of this factor, the computation delay is gradually reduced, but frequent migration leads to excessive migration delay. In contrast, the \textit{JSR} is superior to the above methods, but it may fall into the local optimum due to improper resource allocation. The \textit{DDPG} and \textit{SR-CL} are capable of performing appropriate service migration. In particular, the proposed \textit{SR-CL} introduces the convex optimization theory to achieve optimal resource allocation, thereby obtaining the best performance. Specifically, compared to the \textit{DDPG}, \textit{JSR}, \textit{IDQN}, \textit{GA}, \textit{NM}, and \textit{AM}, the \textit{SR-CL} achieves the performance improvement by around 3.63\%, 9.90\%, 30.96\%, 15.95\%, 20.29\%, and 26.49\%, respectively.

\begin{figure*}
	\centering
    	\begin{minipage}[t]{0.33\linewidth}
		\centering
		\includegraphics[width=2.44in]{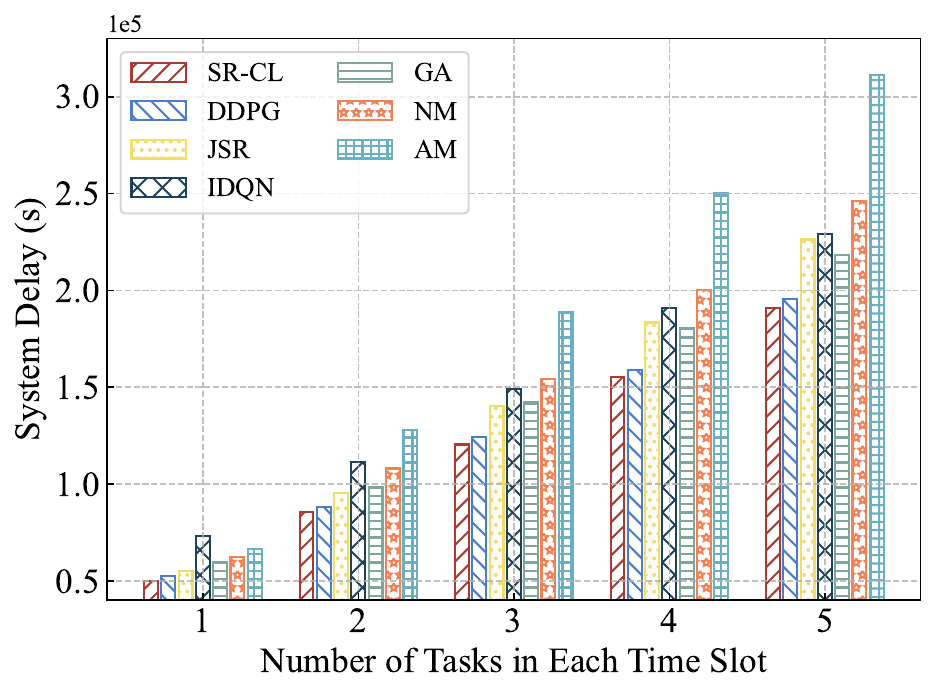}
            \vspace{-0.7cm}
		\caption{Comparison of different methods with dynamic traffic changes.}
		\label{Task_Number}
	\end{minipage}
	\begin{minipage}[t]{0.33\linewidth}
		\centering
		\includegraphics[width=2.44in]{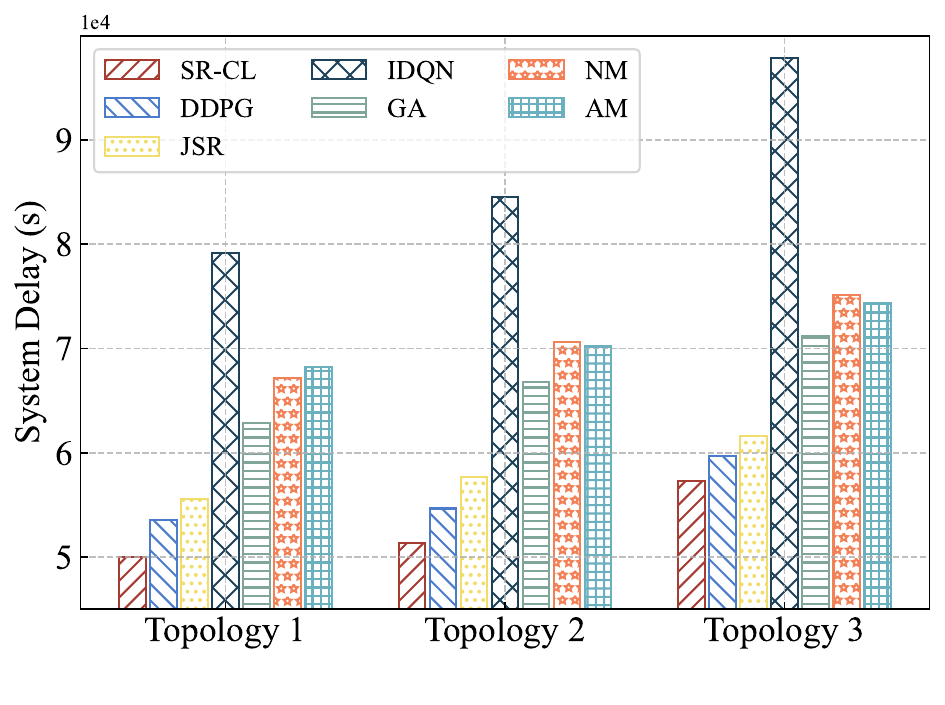}
            \vspace{-0.7cm}
		\caption{Comparison of different methods with various network topologies.}
		\label{Network_Topology}
	\end{minipage}
    \begin{minipage}[t]{0.33\linewidth}
		\centering
		\includegraphics[width=2.44in]{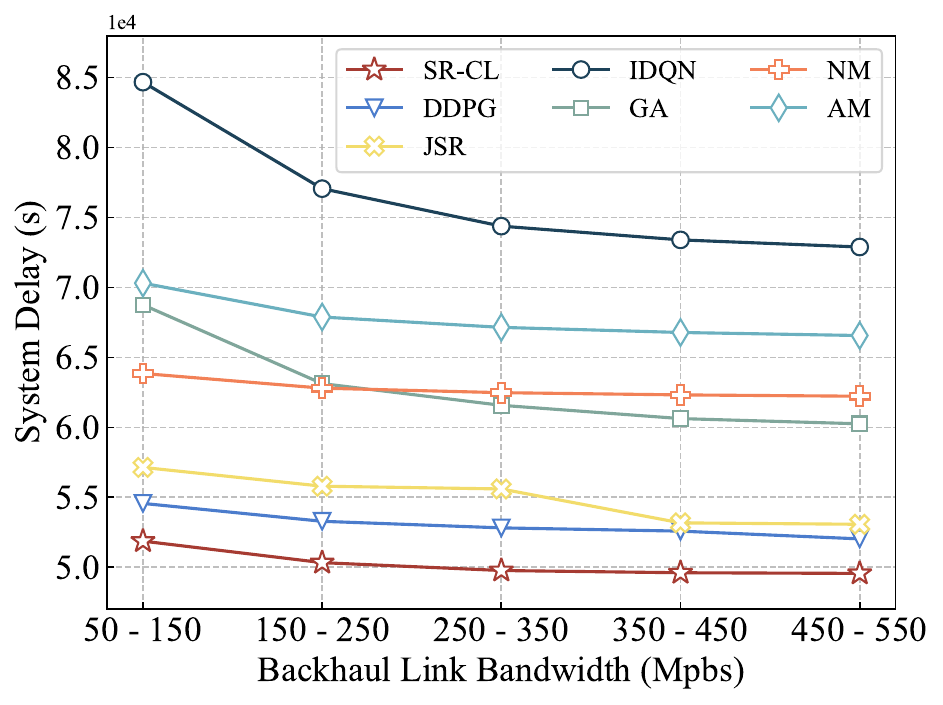}
            \vspace{-0.7cm}
		\caption{Comparison of different methods with various backhaul link bandwidths.}
		\label{Network_Bandwidth}
	\end{minipage}
\end{figure*}

\textbf{Total Delay with Dynamic Traffic Changes.} We evaluate the impact of traffic changes on system delay. Specifically, we simulate traffic changes by varying the number of tasks in each time slot. As shown in Fig. \ref{Task_Number}, system delay increases with rising traffic. This is because the growing number of tasks intensifies the resource competition. The \textit{NM} and \textit{AM} perform well in low-traffic scenarios, but their migration strategies cannot adapt to dynamic traffic changes. The \textit{JSR}, \textit{DDPG}, \textit{GA}, and \textit{IDQN} can adjust their migration strategies through real-time feedback, outperforming the \textit{NM} and \textit{AM}. The \textit{JSR} jointly optimizes service migration and resource allocation but struggles to find the optimal joint strategy due to the large action space. In contrast, the proposed \textit{SR-CL} decouples the sub-problem of resource allocation by introducing the convex-optimization theory and dynamically adjusts the migration strategy via the real-time feedback mechanism of DRL. Therefore, the \textit{SR-CL} can effectively fit into dynamic traffic changes. Specifically, compared to the \textit{DDPG}, \textit{JSR}, \textit{IDQN}, \textit{GA}, \textit{NM}, and \textit{AM}, the \textit{SR-CL} achieves performance improvement of approximately 2.77\%, 14.09\%, 20.11\%, 13.86\%, 21.98\%, and 36.15\%, respectively.

\textbf{Total Delay with Various Unit Migration Delay Coefficients.}
The unit migration delay coefficient is an essential factor that affects the performance of service migration. 
As illustrated in Fig. \ref{Migration_hop}, when there is no service migration by using the \textit{NM}, the change of the coefficient does not influence the system's total delay. With the increase of the coefficient, the performance of all methods (except the \textit{NM}) exhibits a descending trend. When the value of the coefficient grows to 4.5, both the \textit{GA} and \textit{IDQN} become inferior to the \textit{NM} and \textit{AM}. Moreover, the \textit{IDQN} performs worst because it lacks communication between agents and thus cannot capture the complexity and dynamics of the system. In contrast, the proposed \textit{SR-CL} designs the critic with the one-step update to guide the actor with the delayed update for accurately evaluating the Q-values of migration actions. Therefore, the \textit{SR-CL} can always make proper migration decisions to optimize the long-term system delay and outperform the \textit{DDPG} and \textit{JSR}. Especially, when the value of the coefficient is 4.5, the \textit{SR-CL} can reach the performance improvement by approximately 3.62\%, 11.42\%, 54.31\%, 29.29\%, 16.42\%, and 28.59\% compared to the \textit{DDPG}, \textit{JSR}, \textit{IDQN}, \textit{GA}, \textit{NM}, and \textit{AM}, respectively.

\textbf{Total Delay with Various Network Topologies.} As shown in in Fig. \ref{Topologies}, to further evaluate the generality of the proposed \textit{SR-CL} for different network topologies, we construct 3 different topologies of a large-scale network with 25 edge nodes as follows.
\begin{itemize}
    \item Topology 1 (high-connectivity): Each edge node can communicate with its adjacent edge nodes, and thus it is a high-connectivity topology. 
    \item Topology 2 (middle-connectivity): The edge nodes in the central region are with a high-connectivity topology, but the connectivity between edge nodes in the other region is sparse, simulating the typical network topology in cities. 
    \item Topology 3 (low-connectivity): The connectivity between edge nodes is sparse and there is no closed loop, simulating the network topology in remote areas.
\end{itemize}

Fig. \ref{Network_Topology} illustrates the total delay of different methods with various network topologies. As the connectivity of a network topology decreases, the performance of all methods exhibits a downward trend. This is because service migration and task transmission need to be routed via more hops, leading to more time consumption. The \textit{IDQN} lacks efficient collaboration between agents, and thus the decrease in connectivity causes great performance degradation. Since the \textit{AM} and \textit{NM} rely on preset rules, and the decrease in connectivity increases the routing hops of service migration and task transmission. The \textit{GA} employs heuristics with multiple iterations, outperforming the above methods. The \textit{SR-CL}, \textit{DDPG}, and \textit{JSR} adapt to changes in network topologies by interacting with environments. The proposed \textit{SR-CL} utilizes the convex optimization theory to obtain the optimal resource allocation, achieving better performance than the \textit{DDPG} and \textit{JSR} under similar migration decisions. Specifically, compared to the \textit{DDPG}, \textit{JSR}, \textit{IDQN}, \textit{GA}, \textit{NM}, and \textit{AM}, the \textit{SR-CL} achieves the performance improvement by approximately 5.53\%, 9.29\%, 39.35\%, 21.03\%, 25.50\%, and 25.46\%, respectively.

\textbf{Total Delay with Various Backhaul Link Bandwidths.} We evaluate the impact of varying backhaul link bandwidths on the performance of different methods. As shown in Fig. \ref{Network_Bandwidth}, the performance of all methods exhibits an upward trend as the bandwidth increases. This is because high bandwidth reduces the time of transmitting data on the backhaul link, where services instances own more data volumes than tasks and thus consume more transmission time. The \textit{IDQN} fails to capture system dynamics, resulting in frequent service migrations. The \textit{GA} makes migration decisions by randomly searching initialized populations, which also leads to high migration frequency. In the scenarios with low bandwidth, the \textit{GA} performs worse than the \textit{NM}. Compared to the above methods, the \textit{SR-CL}, \textit{DDPG}, and \textit{JSR} can better adapt to the changes of backhaul link bandwidths and avoid unnecessary service migrations. The \textit{DDPG} allocates computational resources in proportion, while the \textit{SR-CL} adopts the convex optimization theory for resource allocation that can further improve system performance. Specifically, compared to the \textit{DDPG}, \textit{JSR}, \textit{IDQN}, \textit{GA}, \textit{NM}, and \textit{AM}, the \textit{SR-CL} achieves the performance improvement by about 5.33\%, 8.61\%, 34.34\%, 20.11\%, 19.94\%, and 25.86\%, respectively.

\begin{table}[!ht] 
	\renewcommand{\arraystretch}{1.3}
	\caption{Average task response delay (s) with various numbers of IVs}
	\label{vehicle_numbers}
	\centering
	\begin{tabular}{|c||c|c|c|c|c|}
		\hline
		\multirow {2}*{\shortstack{\textbf{Methods}}} & 
		\multicolumn{5}{c|}{\textbf{Numbers of IVs}} \\
		\cline{2-6}
		& \tabincell{c}
		{\textbf{60}} &  
		\tabincell{c} 
		{\textbf{100}} &
		\tabincell{c}
		{\textbf{140}} &  
		\tabincell{c} 
		{\textbf{180}} &
		\tabincell{c} 
		{\textbf{220}}\\
		\hline\hline
		\textbf{\textit{SR-CL}} & \textbf{1.5302} & \textbf{2.0876} & \textbf{2.6381} & \textbf{3.1579}  & \textbf{3.7279} \\
		\hline
		\textbf{\textit{DDPG}}  & 1.5871 & 2.1991 & 2.7106 & 3.2505  & 3.8374\\
		\hline
		\textbf{\textit{JSR}} & 1.5778 & 2.3093 & 2.8611 & 3.4857  & 4.2280\\
		\hline
		\textbf{\textit{IDQN}} & 2.4014 & 3.0519 & 3.4952 & 4.2332   & 4.8808\\
		\hline
		\textbf{\textit{GA}} & 2.1002 & 2.4895 & 2.9957 & 3.4720   & 4.0633\\
		\hline
		\textbf{\textit{NM}} & 1.9002 & 2.6032 & 3.1612 & 3.7369   & 4.3681\\
		\hline
		\textbf{\textit{AM}} & 1.8455 & 2.7887 & 3.7321 & 4.5315   & 5.6026\\
		\hline
	\end{tabular}
\end{table}

\textbf{Average Task Response Delay with Various Numbers of IVs.}
As shown in Table \ref{vehicle_numbers}, when the number of IVs is small, MEC servers possess sufficient resources that can be allocated to service instances, and thus there is no obvious performance gap among different methods. 
As the number of IVs grows, it becomes imperative to properly migrate service instances for avoiding high computation delay caused by the overloads of MEC servers. 
Additionally, migration policy also need to consider the high mobility of vehicles to mitigate high communication delay.
The \textit{IDQN} lets each agent make migration decisions without communicating with others, causing the overload issue and local optimum. 
Due to incomplete exploration of continuous space, the \textit{JSR} is inferior to the \textit{GA} as the number of IVs increases.
In contrast, the \textit{DDPG} and \textit{SR-CL} consider the interaction among IVs by observing the global state, enabling better migration decisions. Compared to the \textit{DDPG}, the \textit{SR-CL} incorporates the convex optimization theory to obtain the optimal resource allocation on each MEC server, contributing to the improved performance of service migration. 
Especially, when the number of IVs is 220, the \textit{SR-CL} outperforms the \textit{DDPG}, \textit{JSR}, \textit{IDQN}, \textit{GA}, \textit{NM}, and \textit{AM} by around 2.85\%, 11.83\%, 23.62\%, 8.25\%, 14.66\%, and 33.46\%, respectively.


\begin{table}[!ht] 
	\renewcommand{\arraystretch}{1.3}
	\caption{Decision-making time (ms) of different methods with various numbers of IVs}
	\label{decision_time}
	\centering
	\begin{tabular}{|c||c|c|c|c|c|}
		\hline
		\multirow {2}*{\shortstack{\textbf{Methods}}} & 
		\multicolumn{5}{c|}{\textbf{Numbers of IVs}} \\
		\cline{2-6}
		& \tabincell{c}
		{\textbf{60}} &  
		\tabincell{c} 
		{\textbf{100}} &
		\tabincell{c}
		{\textbf{140}} &  
		\tabincell{c} 
		{\textbf{180}} &
		\tabincell{c} 
		{\textbf{220}}\\
		\hline\hline
		\textbf{\textit{SR-CL}} $ (\times 10^{-1})$ & 7.8726 & 7.8999 & 8.0534 & 8.5952  & 8.6426 \\
		\hline
		\textbf{\textit{DDPG}} $ (\times 10^{-1})$ & 7.4958 & 7.7554 & 8.0017 & 8.0030  & 8.1746\\
		\hline
		\textbf{\textit{JSR}} $ (\times 10^{-1})$ & 8.1798 & 8.6549 & 8.8191 & 9.5902  & 9.9233\\
		\hline
		\textbf{\textit{IDQN}} $ (\times 10^{-1})$ & 1.5480 & 1.5108 & 1.5081 & 1.5080   & 1.5173\\
		\hline
		\textbf{\textit{GA}} $ (\times 10^{3})$ & 1.3669 & 2.4292 & 3.4660 & 4.3594   & 5.1198\\
		\hline
		\textbf{\textit{NM}} $ (\times 10^{-4})$ & 4.9978 & 5.2472 & 4.7883 & 5.1375   & 5.0477\\
		\hline
		\textbf{\textit{AM}} $ (\times 10^{-2})$ & 2.8634 & 2.8672 & 2.8357 & 2.7956   & 2.7983\\
		\hline
	\end{tabular}
\end{table}

\textbf{Decision-making Time with Various Numbers of IVs.} As shown in Table \ref{decision_time}, we evaluate the decision-making time of different methods with various numbers of IVs in a large-scale network topology (25 edge nodes). The \textit{GA} searches for the optimal solution by performing mutation and crossover operations on the individuals in the population. When the number of IVs increases, the time of computing the population fitness significantly increases, degrading the decision-making efficiency. The \textit{NM} always maintains the original migration decisions, and the \textit{AM} makes the migration decisions simply based on the distance between IVs and edge nodes. Therefore, there is no obvious change in the decision-making time of these two methods as the number of IVs increases. The DRL-based methods perform training and inference via neural networks, leading to longer decision-making time than the rule-based methods. The \textit{IDQN} makes decisions only based on local observations, and thus the input and output dimensions of each agent are not affected by the number of IVs. Although the \textit{IDQN} exhibits lower decision-making time than the other DRL-based methods, its decision-making performance is unsatisfying, causing excessive delay. The \textit{SR-CL}, \textit{DDPG}, and \textit{JSR} consider the global environment state as input of neural networks, and thus their state dimensions are related to the number of IVs. The \textit{JSR} learns the policy of resource allocation by neural networks, thereby causing longer decision-making time. Compared to other methods, the proposed \textit{SR-CL} is able to achieve significant performance improvement with satisfying decision-making time.

\subsubsection{Ablation Experiments for SR-CL}

\begin{figure}[!ht]
	\centering
	\begin{center}
		\vspace{-0.2cm}
		\includegraphics*[width=0.83\linewidth]{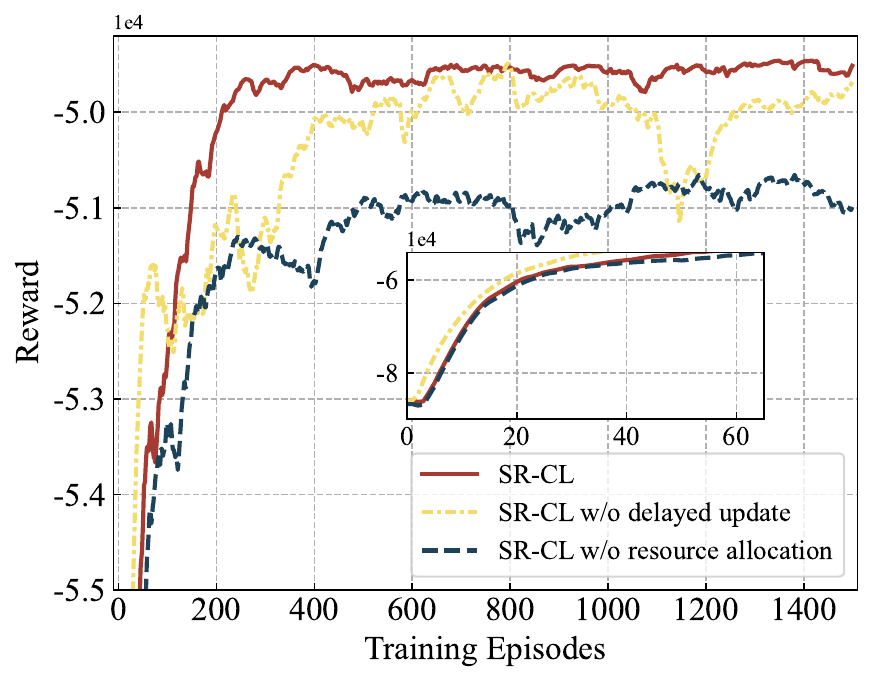}
        \vspace{-0.2cm}
		\caption{Ablation experiments for the proposed \textit{SR-CL}.}
		\vspace{-0.5cm}
        \label{Ablation}
	\end{center}
\end{figure}

We conduct ablation experiments to verify the effectiveness of the components designed in \textit{SR-CL}.
As shown in Fig. \ref{Ablation}, we revert the delayed update to the one-step update, and thus the actor and critic are updated simultaneously, which leads to the high similarity between the weights of the online and target networks. Under this setting, the biased estimation of the target Q-values occurs, which causes inaccurate guidance for the action selection in the actor and seriously impacts the convergence stability of the training process. 
This instability affects the robustness of the \textit{SR-CL} and makes it unsuitable for dynamic multi-edge IoV systems. Furthermore, we modify the convex-optimization-based resource allocation in \textit{SR-CL} to the proportional resource allocation, which allocates resources to service instances in proportion to task demands. However, this manner might not fully optimize the performance of service migration. In contrast, when adopting the convex-optimization-based resource allocation, the \textit{SR-CL} can guarantee the global optimal solution with the support of the mathematical theory. Therefore, the \textit{SR-CL} is able to promise high adaptability in addressing the complex problem of service migration and resource allocation in changeable multi-edge IoV systems.

\subsection{Simu5G-based Testbed Validation}

\begin{figure}[!ht]
	\centering
	\begin{center}
		\vspace{-0.2cm}
		\includegraphics*[width=1.0\linewidth]{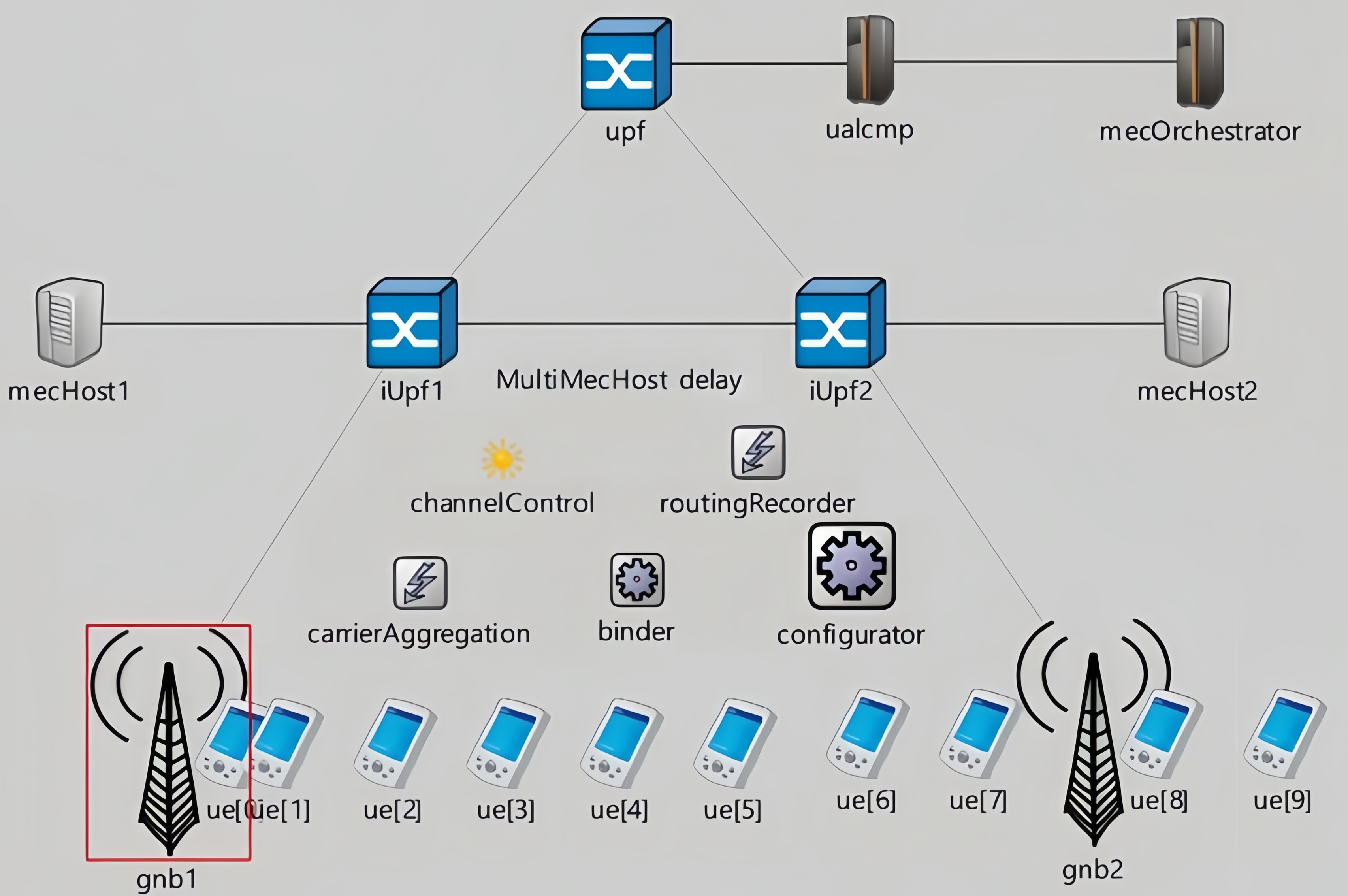}
            \vspace{-0.4cm}
		\caption{Topology of Simu5G-based testbed.}
            \vspace{-0.2cm}
		\label{Simu5G_testbed}
	\end{center}
\end{figure}

Simu5G \cite{Nardini2020Simu} is an open-source framework built on OMNeT++, which is designed to simulate the key features and complex scenarios of 5G networks. Simu5G supports the simulation of modules such as the 5G Radio Access Network (RAN), Core Network (CN), and Multi-access Edge Computing (MEC), which can be used to simulate the communication interactions among User Equipments (UEs), base stations (i.e., \textit{gNodeBs} (\textit{gNBs})), edge nodes, and the remote cloud. Based on a PC with an Intel(R) Core(TM) i5-12400F CPU and 16GB RAM, we install OMNeT++ 6.1.0 and INET 4.5.2 and build a Simu5G-based testbed\footnote{\url{https://github.com/JoyZhang20/SRCL_Simu5G}}. As shown in Fig. \ref{Simu5G_testbed}, the topology of the Simu5G-based testbed includes mobile UEs, \textit{gNBs} that provide wireless access, \textit{mecHost} that offers computing services, User Plane Function (UPF) and Intermediate UPF (IUPF) for transmitting routing data between the access network and edge nodes, User Application Layer Control and Management Plane (UALCMP) for application-level control and management, \textit{mecOrchestrator} for application deployment, resource allocation, and service scheduling, and \textit{channelControl} for channel management.

On the Simu5G-based testbed, we set the transmission power to 26 dBm for UEs and 46 dBm for \textit{gNBs} and turn on the interference in uplink and downlink. We randomly place 10 UEs within the communication coverage areas of 2 \textit{gNBs} with different distances to \textit{gNBs}. All UEs follow the movement pattern of \textit{LinearMobility} with a speed of 5 m/s, \textit{updateInterval} is 0.05 s, and \textit{dynamicCellAssociation }and \textit{enableHandover} are enabled. We regard \textit{requestResponseApp} provided by Simu5G as the service requested by UEs, which is deployed in \textit{mecHost} with \textit{requestServiceTime} of 4 ms and {subscriptionServiceTime} of 11 µs. With the ETSI GS MEC 003 standard, Simu5G uses Virtualization Infrastructure Manager (VIM), a part of \textit{mecOrchestrator}, to allocate, manage, and release virtualized infrastructure resources. VIM has two built-in scheduling strategies (i.e., \textit{SEGREGATION} and \textit{FAIR}). The \textit{SEGREGATION} allocates resources strictly according to predefined demands, and the \textit{FAIR} considers both demands and current loads to allocate resources proportionally. We implement the resource allocation policy based on the convex-optimization theory in \textit{calculateProcessingTime} function of VIM. For service migration, we construct the DRL agent in Python programs, treating Simu5G as the environment and making interactions. Simu5G offers many delay metrics including \textit{downLinkTime}, \textit{upLinkTime}, \textit{serviceResponseTime}, \textit{processingTime}, and \textit{responseTime}, where \textit{responseTime} is the total delay. During the offline training phase, we randomly make service migration decisions and collect delay data after each simulation to train the DRL agent. Due to the limited compatibility of Simu5G with mainstream communication frameworks (e.g., curl and Boost.Asio), we use local files to transfer the decisions of the DRL agent during the real-time evaluation phase and synchronize Python programs and Simu5G. Other parameters are set by using the default configurations of Simu5G.

\begin{figure}[!ht]
	\centering
	\begin{center}
		\vspace{-0.2cm}
		\includegraphics*[width=0.88\linewidth]{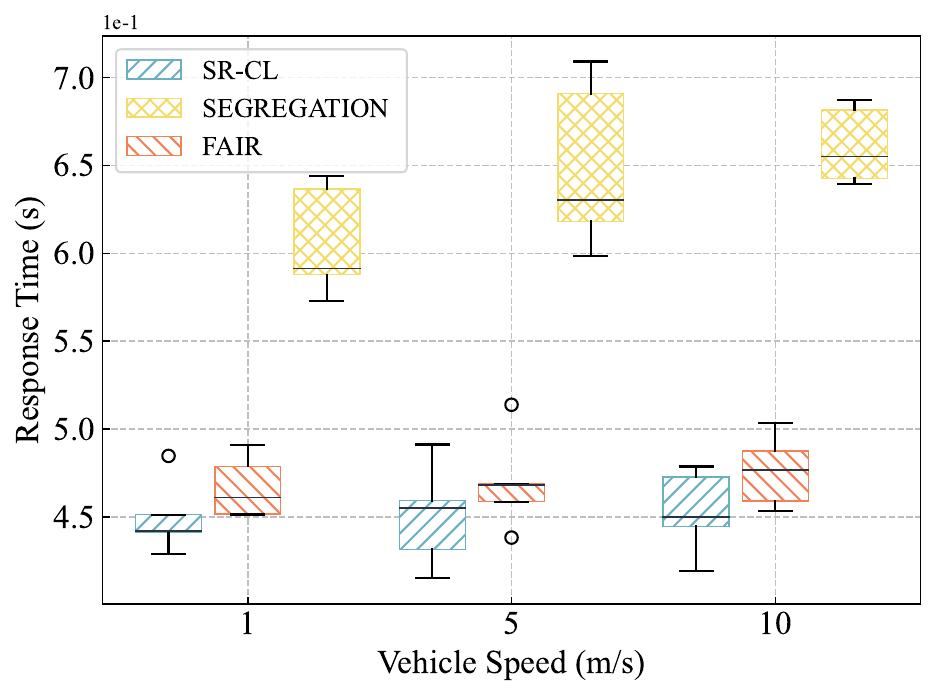}
        \vspace{-0.4cm}
		\caption{Comparison of different methods with various vehicle speeds.}
        \vspace{-0.2cm}
		\label{Vehicle_speed}
	\end{center}
\end{figure}

First, we compare the response time (i.e., \textit{responseTime}) of different methods with various vehicle speeds. As illustrated in Fig. \ref{Vehicle_speed}, the \textit{SEGREGATION} exhibits the highest \textit{responseTime}. This is because the \textit{SEGREGATION} employs static resource allocation but ignores the load of \textit{mecHost}, and thus some tasks cannot be allocated with proper resources. Different from the \textit{SEGREGATION}, the \textit{FAIR} considers the load variations of \textit{mecHost} and allocates resources proportionally, thereby enhancing resource efficiency and reducing \textit{responseTime} to a certain extent. Compared to other methods, the proposed \textit{SR-CL} can allocate resources according to task features, which further improves resource efficiency and \textit{responseTime}. Meanwhile, the \textit{SR-CL} incorporates proactive service migration, which can make a better balance between migration delay and load status, and then it can choose a more reasonable \textit{mecHost} with lower delay for offloading tasks. As the vehicle speed increases, \textit{responseTime} of all methods rises. On one hand, faster movement causes signal instability, increasing both \textit{downLinkTime} and \textit{upLinkTime}. On the other hand, vehicle mobility between different \textit{gNBs} introduces extra \textit{responseTime}. The results demonstrate the superior performance of the \textit{SR-CL} under scenarios with varying vehicle speeds.

\begin{figure}[!ht]
	\centering
	\begin{center}
		\vspace{-0.2cm}
		\includegraphics*[width=0.82\linewidth]{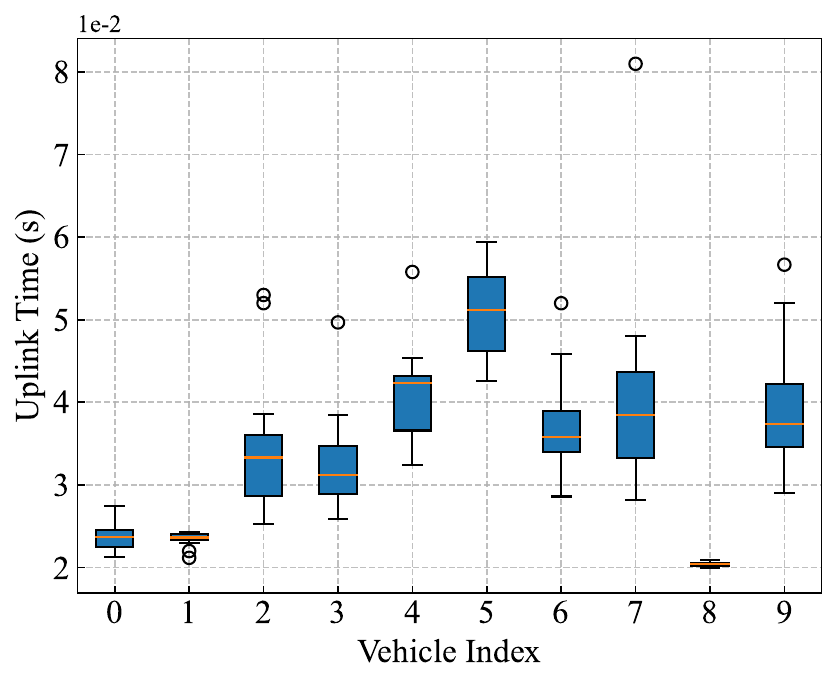}
        \vspace{-0.4cm}
		\caption{Comparison of uplink time on different locations of vehicles.}
            \vspace{-0.5cm}
		\label{Vehicle_index}
	\end{center}
\end{figure}

Next, we compare the uplink time (i.e., \textit{upLinkTime}) on different locations of vehicles. As depicted in Fig. \ref{Vehicle_index}, each box indicates the distribution of \textit{upLinkTime} during 15 offloading processes between vehicles and \textit{mecHost}. \textit{UpLinkTime} is closely related to the locations of vehicles. For example, the vehicles that are closer to \textit{gNBs} experience less signal attenuation and faster upload rates, leading to lower \textit{upLinkTime} and more concentrated distributions. In contrast, the vehicles that are farther from \textit{gNBs} face more severe signal attenuation and more interference during movement, resulting in greater fluctuations in upload rates, higher \textit{upLinkTime}, and more dispersed distributions. When vehicles are located at the boundary of the communication coverage of two \textit{gNBs} (i.e., index = 5), the highest \textit{upLinkTime} happens. This is because vehicles trigger the switching of \textit{gNBs} during movement, which introduces extra communication overheads and further increases \textit{upLinkTime}.

\section{Conclusion}\label{sec:conclusion}
In this paper, we propose \textit{SR-CL}, a novel mobility-aware seamless service migration and resource allocation framework for multi-edge IoV systems. First, we formulate the service migration and resource allocation as a long-term QoS optimization problem and decouple it into two sub-problems. For the sub-problem of service migration, we design a new improved DRL-based method with delayed and one-step update mechanisms. For the sub-problem of resource allocation, we theoretically derive the optimal resource allocation based on convex optimization. Using the real-world datasets of vehicle trajectories and testbed, extensive experiments are conducted to verify the effectiveness of the proposed \textit{SR-CL}. Compared to benchmark methods, the \textit{SR-CL} achieves better performance under different scenarios with various computational capabilities, unit migration delay coefficients, network topologies, backhaul link bandwidths, and numbers of IVs. Meanwhile, the \textit{SR-CL} exhibits faster and more stable convergence than the advanced \textit{IDQN} and \textit{DDPG}. Further, testbed experiments validate the practicality and superiority of the \textit{SR-CL} in offering seamless services. In our future work, we plan to integrate energy costs into our optimization objective and explore a distributed solution to better balance delay and energy costs, aiming to further enhance the scalability of the proposed system in real-world scenarios. Moreover, we will explore more recent datasets of vehicle trajectories from different cities to evaluate the generality of the \textit{SR-CL} in various scenarios and extend the testbed for further approaching real-world environments.

\section*{Acknowledgments}
This work was partly supported by the National Natural Science Foundation of China under Grant No. 62202103, the Central Funds Guiding the Local Science and Technology Development under Grant No. 2022L3004, the National Natural Science Foundation of Chongqing under Grant No. CSTB2024NSCQ-JQX0013, the Fujian Province Technology and Economy Integration Service Platform under Grant No. 2023XRH001, and the Fuzhou-Xiamen-Quanzhou National Independent Innovation Demonstration Zone Collaborative Innovation Platform under Grant No. 2022FX5.

\bibliographystyle{ieeetr}
\bibliography{reference}

\ifCLASSOPTIONcaptionsoff
  \newpage
\fi

\vspace{0.30cm}
\begin{IEEEbiography}[{\includegraphics[width=0.9in,height=1.125in,clip,keepaspectratio]{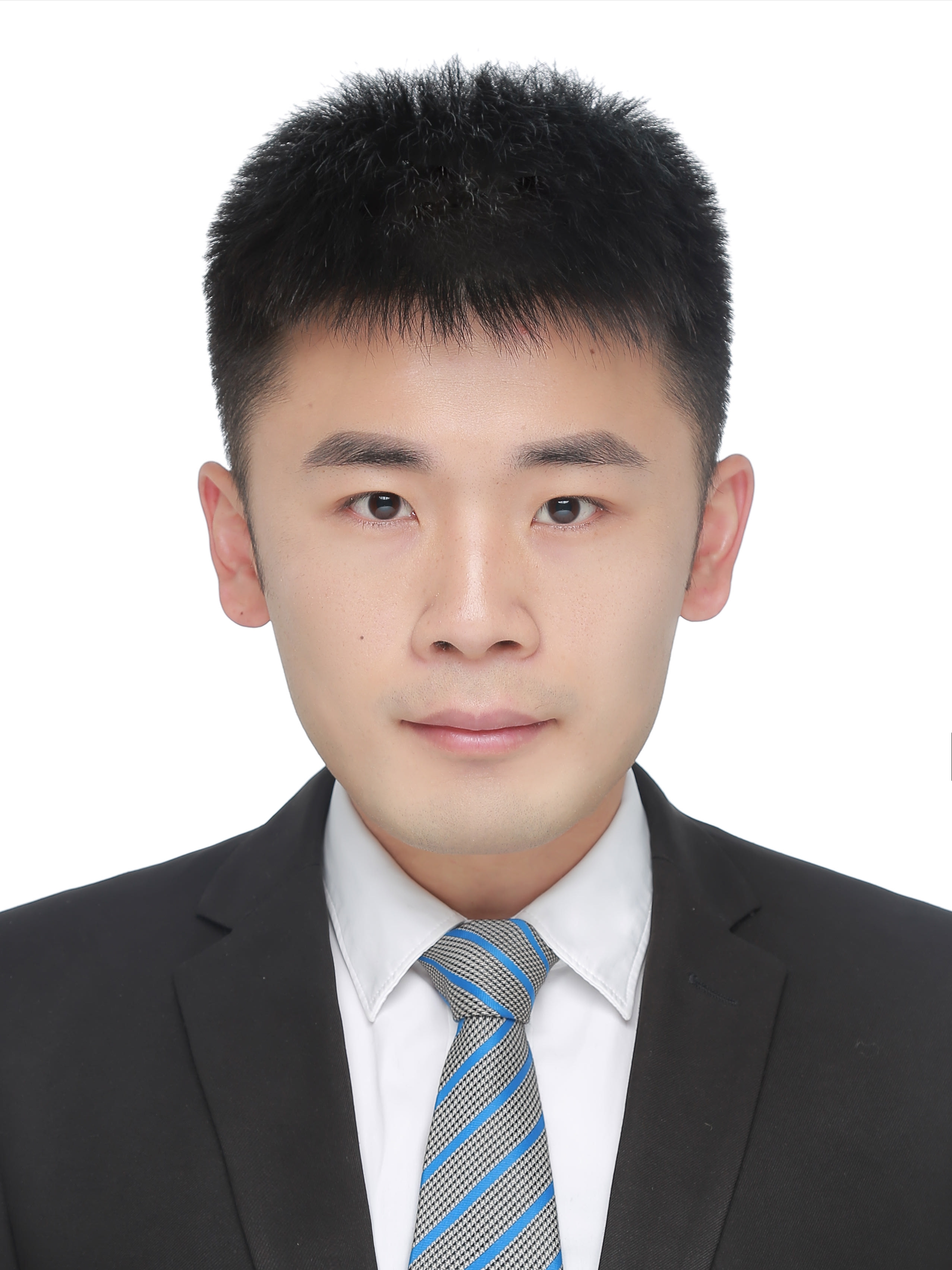}}]{Zheyi Chen}
    is a Professor and Qishan Scholar with the College of Computer and Data Science at the Fuzhou University, China. He received his Ph.D. degree in Computer Science from the University of Exeter, UK, in 2021, and his M.Sc. degree in Computer Science and Technology from the Tsinghua University, China, in 2017, respectively. His research interests include cloud-edge computing, resource optimization, deep learning, and reinforcement learning. Dr. Chen has published over 40 research papers in reputable international journals and conferences such as IEEE TPDS, IEEE JSAC, IEEE TMC, IEEE/ACM ToN, IEEE INFOCOM, ACM SIGKDD, IEEE TII, IEEE ComMag, IEEE Network, IEEE IoTJ, FGCS, IEEE TCC, and IEEE ICC.

\end{IEEEbiography}

\begin{IEEEbiography}[{\includegraphics[width=0.9in,height=1.125in,clip,keepaspectratio]{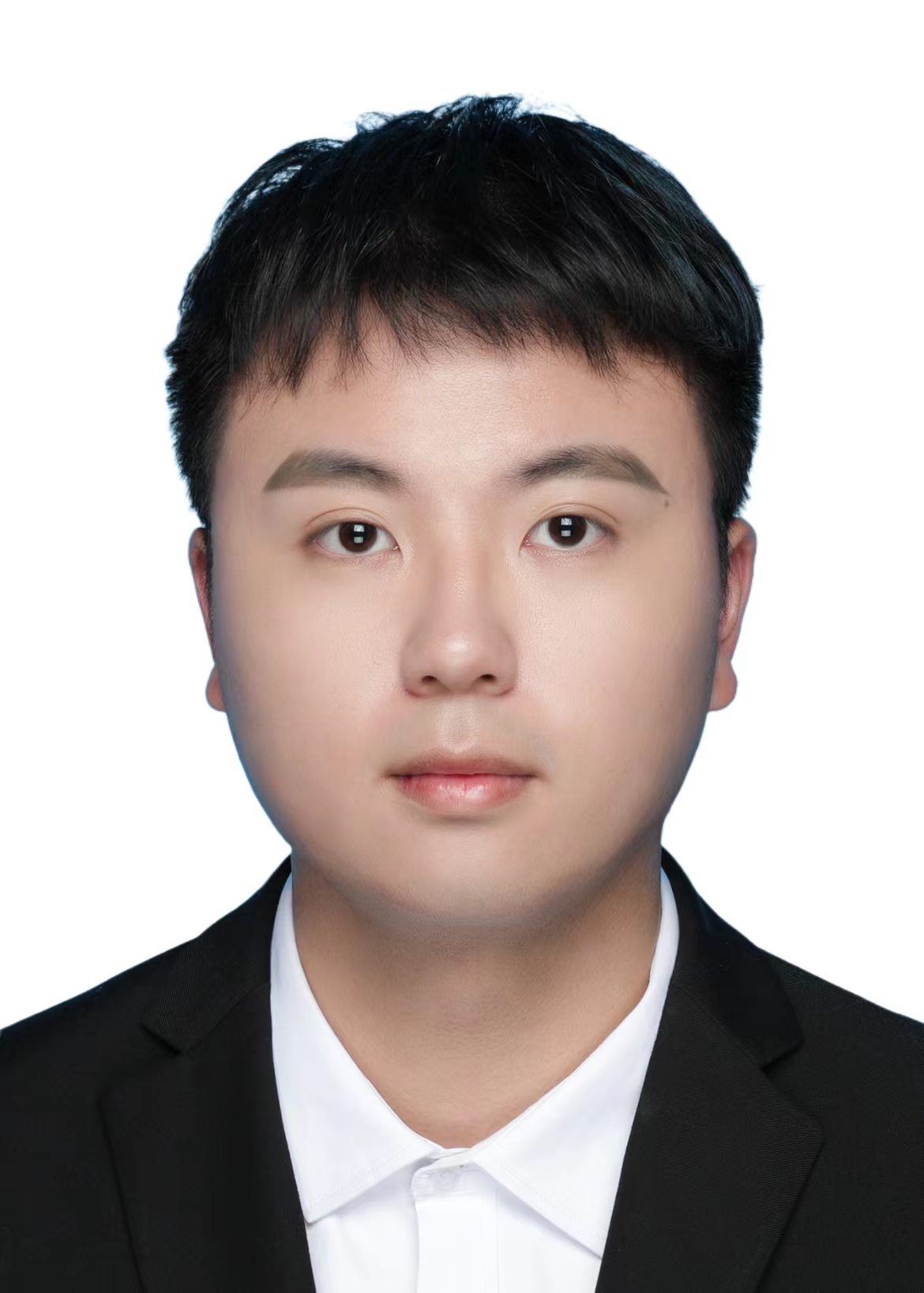}}]{Sijin Huang}
	is currently working toward his M.S. degree in Computer Applied Technology with the College of Computer and Data Science at the Fuzhou University. He received his B.S. degree in Computer Science and Technology from Huaqiao University, China, in 2022. His current research interests include cloud/edge computing, deep reinforcement learning and service migration.

\end{IEEEbiography}


%

\begin{IEEEbiography}[{\includegraphics[width=0.9in,height=1.125in,clip,keepaspectratio]{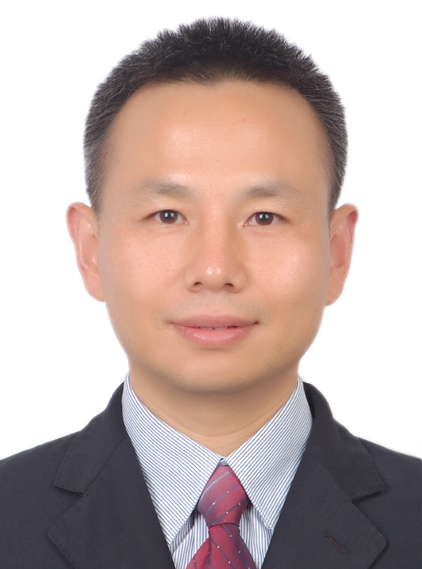}}]{Geyong Min}
	is a Professor of High Performance Computing and Networking in the Department of Computer Science within the Faculty of Environment, Science and Economy at the University of Exeter, United Kingdom. He received the Ph.D. degree in Computing Science from the University of Glasgow, United Kingdom, in 2003, and the B.Sc. degree in Computer Science from Huazhong University of Science and Technology, China, in 1995. His research interests include future Internet, computer networks, wireless communications, multimedia systems, information security, high-performance computing, ubiquitous computing, modelling, and performance engineering.

\end{IEEEbiography}
\vfill
\newpage

\begin{IEEEbiography}[{\includegraphics[width=0.9in,height=1.125in,clip,keepaspectratio]{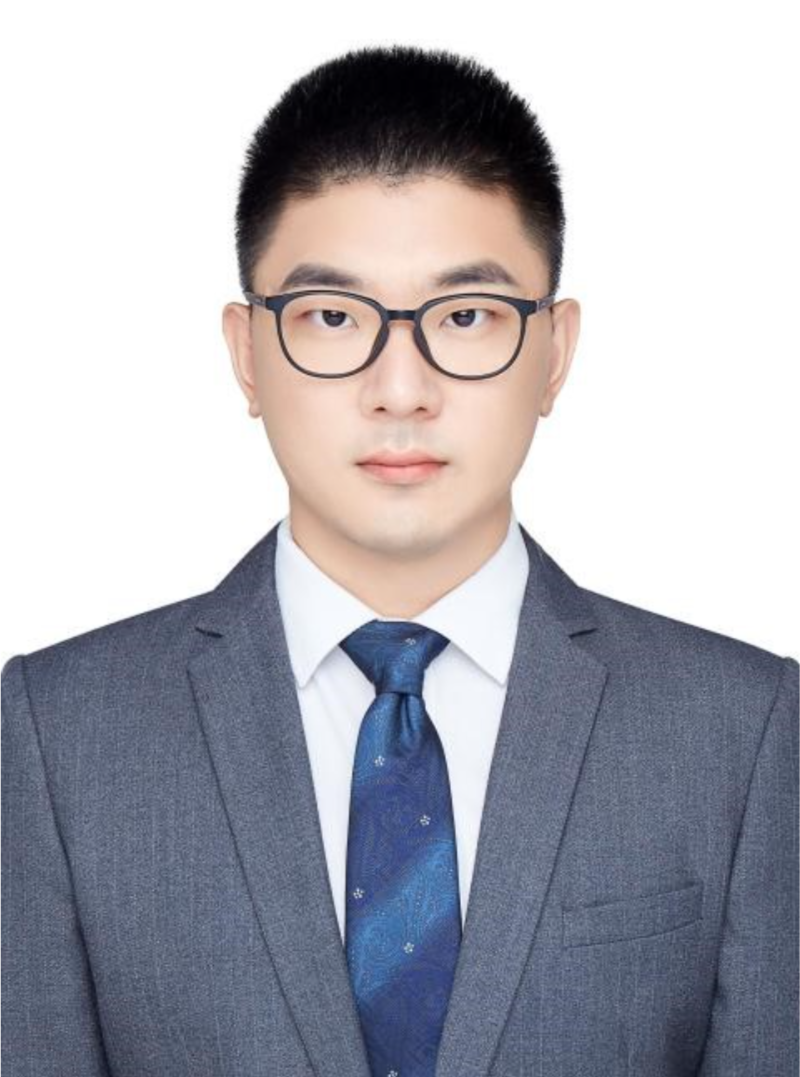}}]{Zhonglong Ning}
	(Senior Member, IEEE) received the Ph.D. degree from Northeastern University, China in 2014. He was a Research Fellow at Kyushu University from 2013 to 2014, Japan. Currently, he is a full professor with the School of Communications and Information Engineering, Chongqing University of Posts and Telecommunications, Chongqing, China. His research interests include mobile edge computing, 6G networks, machine learning, and resource management. He has published over 150 scientific papers in international journals and conferences. Dr. Ning serves as an associate editor or guest editor of several journals, such as IEEE Transactions on Industrial Informatics, IEEE Transactions on Social Computational Systems, IEEE Internet of Things Journal and so on. He is a Highly Cited Researcher (Web of Science) since 2020.

\end{IEEEbiography}

\begin{IEEEbiography}[{\includegraphics[width=0.9in,height=1.125in,clip,keepaspectratio]{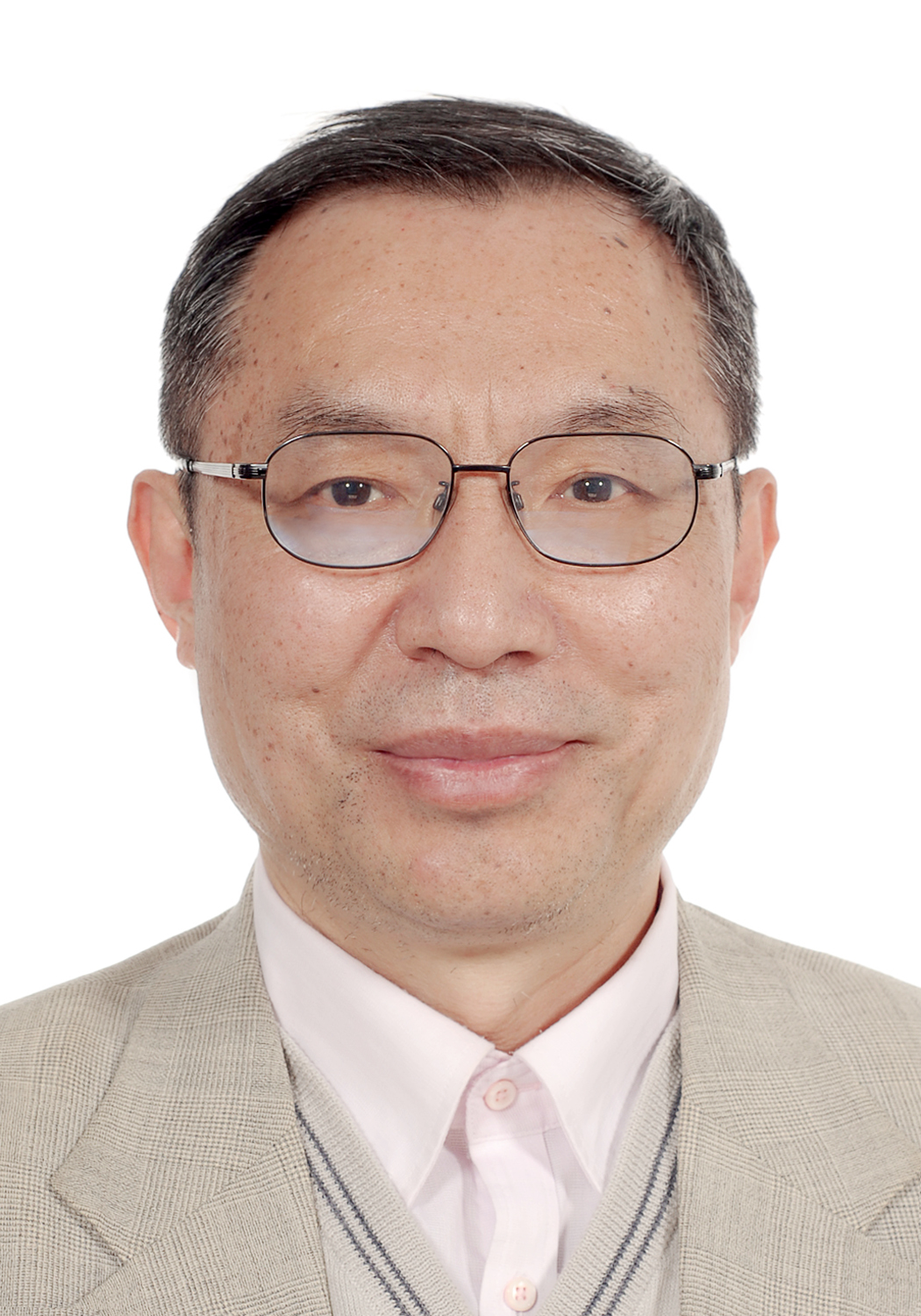}}]{Jie Li} 
	(Fellow, IEEE) received the B.E. degree in computer science from Zhejiang University, Hangzhou, China, the M.E. degree in electronic engineering and communication systems from China Academy of Posts and Telecommunications, Beijing, China. He received the Dr. Eng. degree from the University of Electro-Communications, Tokyo, Japan. He is with Department of Computer Science and Engineering, Shanghai Jiao Tong University, Shanghai, China where he is a chair professor. His current research interests are in big data and AI, blockchain, edge computing, networking and security, OS, information system architecture. He serves as the director of Shanghai Jiao Tong University Blockchain Research Centre. He was a professor in Department of Computer Science, University of Tsukuba, Japan. He was a visiting Professor in Yale University, USA, Inria Sophia Antipolis and Inria Grenoble-Rhone-Aples, France. He is the co-chair of IEEE Technical Community on Big Data and the founding Chair of IEEE ComSoc Technical Committee on Big Data. He serves as an associated editor for many IEEE journals and transactions. He has also served on the program committees for several international conferences.

\end{IEEEbiography}

\begin{IEEEbiography}[{\includegraphics[width=0.9in,height=1.125in,clip,keepaspectratio]{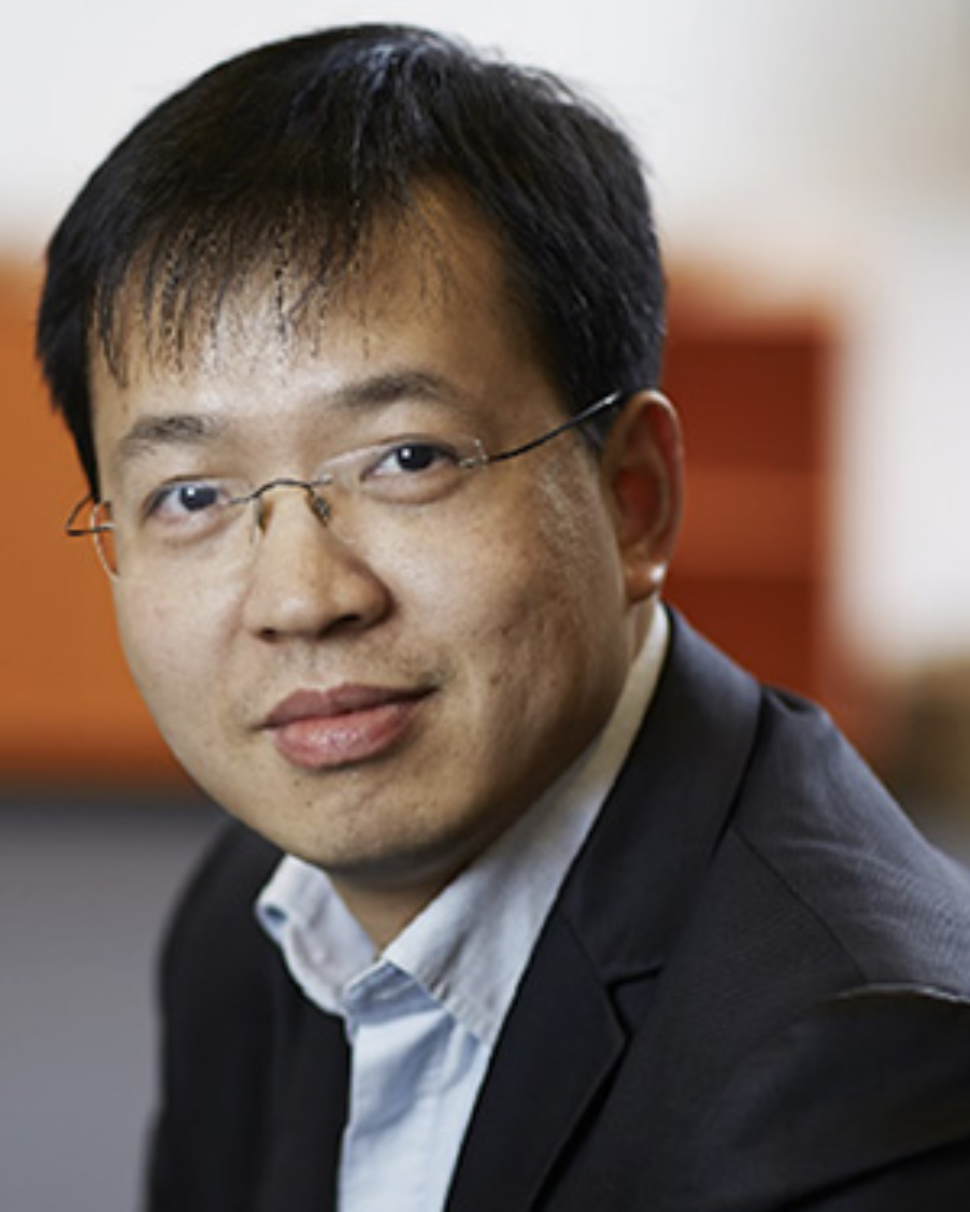}}]{Yan Zhang} 
	(Fellow, IEEE) received the Ph.D. degree from the School of Electrical and Electronics Engineering, Nanyang Technological University, Singapore. He is currently a Full Professor with the Department of Informatics, University of Oslo, Oslo, Norway. His research interests include next-generation wireless networks leading to 6G and green and secure cyber-physical systems (e.g., smart grid and transport). He was a recipient of the Global Highly Cited Researcher Award (Web of Science top 1\% most cited worldwide), since 2018. He is a Symposium/Track Chair in a number of conferences, including IEEE ICC 2021, IEEE Globecom 2017, IEEE PIMRC 2016, and IEEE SmartGridComm 2015. He is the Chair of IEEE Communications Society Technical Committee on Green Communications and Computing. He is an Editor (or an Area Editor, a Senior Editor, and an Associate Editor) for several IEEE Transactions/magazines, including IEEE Communications Magazine, IEEE Network Magazine, IEEE Transactions on Network Science and Engineering, IEEE Transactions on Vehicular Technology, IEEE Transactions on Industrial Informatics, IEEE Transactions on Green Communications and Networking, IEEE Communications Survey and Tutorials, IEEE Internet of Things Journal, IEEE Systems Journal, IEEE Vehicular Technology Magazine, and IEEE Blockchain Technical Briefs. He is a CCF Senior Member, an Elected Member of CCF Technical Committee of Blockchain, and a CCF Distinguished Speaker in 2019. He is a Fellow of IET and an Elected Member of Academia Europaea and the Norwegian Academy of Technological Sciences.

\end{IEEEbiography}
\vfill

\end{document}